\documentclass[useAMS,usenatbib]{mn2e}
\usepackage{graphicx,color}

\usepackage{times,ulem}

\newcommand{\alm}{a_{\ell m}}
\newcommand{\Cl}{C_\ell}
\newcommand{\ellrecon}{\ell_{\mathrm{recon}}}
\newcommand{\ellmax}  {\ell_{\mathrm{max}}}
\newcommand{\Npix}{N_{\mathrm{pix}}}

\newcommand{\trace}{\mathop{\mathrm{Tr}}\nolimits}
\newcommand{\transpose}[1]{{#1}^{\mathsf{T}}}
\newcommand{\mat}[1]{\mathbfss{#1}}
\newcommand{\Cinv}{\mat{C}^{-1}}
\newcommand{\nside}{\textsc{Nside}}
\newcommand{\muK}{\mathrm{\umu K}}
\newcommand{\spice}{\textsc{SpICE}}
\newcommand{\healpix}{\textsc{HEALPix}}

\newcommand{\satellite}[1]{\textit{#1}}
\newcommand{\COBE}{\satellite{COBE}}
\newcommand{\WMAP}{\satellite{WMAP}}

\newcommand{\unit}[1]{\;\mathrm{#1}}
\renewcommand{\vec}[1]{\bmath{#1}}
\newcommand{\unitvec}[1]{\vec{\hat{#1}}}

\newcommand{\dderiv}{\mathrm{d}}

\title[CMB Reconstruction bias]{Bias in low-multipole CMB reconstructions}
\author[C.J. Copi, D. Huterer, D.J. Schwarz and G.D. Starkman]
{Craig J. Copi$^{1}$\thanks{E-mail: cjc5@cwru.edu},
  Dragan Huterer$^{2}$\thanks{E-mail: huterer@umich.edu},
  Dominik J. Schwarz$^{3}$\thanks{E-mail: dschwarz@physik.uni-bielefeld.de}
  and 
  Glenn D. Starkman$^{1}$\thanks{E-mail: glenn.starkman@case.edu}\\
  $^{1}$CERCA/Department of Physics/ISO, Case Western Reserve University, Cleveland, 
  OH 44106-7079, USA\\
  $^{2}$Department of Physics, University of Michigan, 
  450 Church St, Ann Arbor, MI 48109-1040, USA\\
  $^{3}$Fakult\"at f\"ur Physik, Universit\"at Bielefeld, Postfach 100131, 
  33501 Bielefeld, Germany}

\begin{document}

\date{Accepted xxxx. Received xxxx; in original form xxxx}

\pagerange{\pageref{firstpage}--\pageref{lastpage}} \pubyear{2011}

\maketitle

\label{firstpage}

\begin{abstract}
   The large-angle, low multipole cosmic microwave background~(CMB)
   provides a unique view of the largest angular scales in the
   Universe. Study of these scales is hampered by the facts that we have
   only one Universe to observe, only a few independent samples of the
   underlying statistical distribution of these modes, and an incomplete
   sky to observe due to the interposing Galaxy.  Techniques for
   reconstructing a full sky from partial sky data are well known and have
   been applied to the large angular scales.  In this work we critically
   study the reconstruction process and show that, in practise, the
   reconstruction is biased due to leakage of information from the region
   obscured by foregrounds to the region used for the reconstruction.  We
   conclude that, despite being suboptimal in a technical sense, using the
   unobscured region without reconstructing is the most robust measure of
   the true CMB sky. We also show that for noise free data reconstructing
   using the usual optimal, unbiased estimator may be employed without
   smoothing thus avoiding the leakage problem.  Unfortunately, directly
   applying this to real data with noise and residual, unmasked foregrounds
   yields highly biased reconstructions requiring further care to apply
   this method successfully to real-world CMB.
\end{abstract}

\begin{keywords}
cosmology: cosmic microwave background --
cosmology: large-scale structure of Universe
\end{keywords}

\section{Introduction}

Several prominent anomalies in the large-angle, low-$\ell$ cosmic microwave
background (CMB) have been identified, starting with pioneering
observations by the \satellite{Cosmic Background
  Explorer}~(\COBE)~\citep{DMR4}, and confirmed and extended with the high
precision observations from the \satellite{Wilkinson Microwave Anisotropy
  Probe}~(\WMAP)~\citep{WMAP1-maps}. These anomalies include the
unexpectedly low correlations at scales above 60 degrees \citep{DMR4,
  WMAP1-maps,CHSS-review,SHCSS2011},the alignments of the largest
multipoles with each other and the Solar System
\citep{deOliveira-Costa2004,SSHC2004,Land-MPV,CHSS-anomalies}, a parity
asymmetry at low multipoles \citep{KN2010a,KN2010b,KN2010c,KN2010d}, and
the spatial asymmetries in the distribution of power observed at smaller
scales~\citep{EHBGL2004, EHBGL2004Erratum, Hansen2009}.  Numerous attempts
have been made to explain or explain away these anomalies
\citep{Slosar2004,Hajian2007, Afshordi-extradim,WMAP7-anomalies} -- none of
them successful \citep[see][and references therein, for a
  review]{CHSS-review}.

The most peculiar and robust CMB anomaly is arguably the lack of correlation on
large angular scales first observed by \COBE~\citep{DMR4} and confirmed and
further quantified through the $S_{1/2}$ statistic by
\WMAP~\citep{WMAP1-cosmology}. Subsequent study of the two point angular
correlation function, $C(\theta)$, has found further oddities; the large
angle correlation is mainly missing outside of the Galactic region, there
being essentially no correlation on large angles.  The large-angle
correlation that is observed comes from the foreground removed Galactic
region of the reconstructed full-sky map~\citep{CHSS-WMAP5}.  From the
internal linear combination (ILC) map,\footnote{The ILC map and all data
  from the \WMAP{} mission is freely available on-line at
  \texttt{http://lambda.gsfc.nasa.gov/}.} the full-sky map created from the
individual frequency bands which provides our best picture of the full
sky microwave background radiation, it is found that the lack of
correlation is unlikely at the approximately $95$ per cent level.  However,
when solely the region outside the Galaxy of the individual frequency or
ILC maps are analysed the lack of correlation is rare at the approximately
$99.975$ percent level~\citep{CHSS-WMAP5}.

The study of the large-angle CMB presents special problems that must be
treated carefully.  Since there is only one Universe to observe and few
independent modes at low-$\ell$, large sky coverage is needed, and even with
this coverage, very little independent information about the ensemble is available.
Further, given the observed low quadrupole power, $C_2^{\mathrm obs}\sim
100$--$200\unit{(\muK)^2}$, compared to the best fit $\Lambda$CDM model,
$C_2^{\Lambda\mathrm{CDM}}\sim1300\unit{(\muK)^2}$, large-angle studies are
particularly sensitive to assumptions and unintended biases.

One suggestive example of this is provided by the ILC map itself.  If we
use a pixel based estimator for the $\Cl$ as implemented in
\spice~\citep{polspice} we can easily determine the quadrupole power inside
and outside the \WMAP{} provided analysis mask KQ75y7 to be
\begin{equation}
  C_2^{\mathrm{inside}} \approx 610\unit{(\muK)^2}, \qquad
  C_2^{\mathrm{outside}} \approx 80\unit{(\muK)^2}.
\end{equation}
The KQ75y7 mask cuts out approximately $25$ percent of the sky.  Taking the
weighted average of these values produces the intriguing result
\begin{equation}
  0.25 C_2^{\mathrm{inside}} + 0.75 C_2^{\mathrm{outside}} \approx
  200\unit{(\muK)^2},
\end{equation}
a value consistent with the \WMAP{} reported
$C_2$~\citep{WMAP7-angular-ps}.  Again we stress this is a suggestive
example, not a careful analysis; the pseudo-$\Cl$ (PCL) estimator employed
here is suboptimal, we have not include errors on the estimates, etc.  It
does, however, show the wide discrepancy between the Galactic region and
the rest of the sky, a common theme for the ILC map.  Further it shows how
a large value mixed in from a small region of the sky significantly impacts
the final result.

In a recent paper \cite{Efstathiou2010}, the authors claimed that the
low $S_{1/2}$ results are due to the use of a suboptimal estimator
(the pixel based estimator) of $C(\theta)$ and proposed an alternative
based on reconstructing the full sky.  This proposal avoids addressing
the question of \textit{why} the partial sky contains essentially no
correlations on large angular scales and instead focuses on a new
question that centre on the issue of how the full sky is reconstructed. In
this work we carefully study full-sky reconstruction algorithms and
their effects on the low-$\ell$ CMB\@.

It is well known that contamination affects the reconstruction of the low
multipoles \citep{Bielewicz2004,Naselsky2008,Liu2009,Aurich2010}.  In
particular \cite{Aurich2010} have found that smoothing of full sky map
prior to analysis, as required by a reconstruction algorithm (see
\cite{Efstathiou2010} and our discussion below) leaks information from from
the region inside the mask to pixels outside the mask.  They showed that
the pixels outside the mask have errors that are a significant fraction of
the mean CMB temperature.  They further find that it is safest to calculate
the two point angular correlation function on the cut-sky.  Here we confirm
and extend these results.

Alternative analyses such as that suggested in \cite{Efstathiou2010}, must
be performed with care.  In this work we carefully study the full-sky
reconstruction, based on the cut-sky data, in a Universe with low
quadrupole power.  In Sec.~\ref{sec:formalism} we briefly present the
formalism typically employed in CMB studies.  Sec.~\ref{sec:results}
contains our results and we conclude in Sec.~\ref{sec:conclusions}.
Ultimately we find that if a full-sky map, such as the ILC, is a faithful
representation of the true CMB sky, then a reconstruction algorithm can
reproduce its properties.  This is not surprising: if the full-sky map is
already trusted, there is no need to perform a reconstruction and nothing
is gained by doing so.  However, if part of the full sky is not trusted or is
known to be contaminated, then, by reconstructing without properly
accounting for  the assumptions implicit in the algorithm, the final results
will be biased toward the full-sky values.  Again this is not surprising,
if information from the questioned region is allowed to leak into the rest
of the map then it will affect the final results and nothing will be
learned about the validity of the reconstruction.

In any reconstruction of unknown values from the properties of existing
data assumptions must be made.  Often these assumptions are not explicitly
stated.  For the work presented here we take the observed microwave sky
outside of the Galactic region as defined through the KQ75y7 mask to be a
fair sample of the CMB\@.  This partial-sky region is known to have
essentially no correlations on large angular scales; it is unlikely in the
best fit $\Lambda$CDM model at the $99.975$ per cent level \citep{CHSS-WMAP5}.
Our study shows the bias introduced into full-sky reconstructions when
an admixture of a region with larger angular correlations is included prior
to reconstruction.  We stress that results of the partial-sky analysis are
\textit{not} being questioned, instead a new question is being asked; how
should the full sky be reconstructed when there is a wide disparity between
the statistical properties of the region outside the Galaxy and that inside.

\section{Reconstruction Formalism}
\label{sec:formalism}

Optimal, unbiased estimators for both the $\Cl$ and $\alm$ are well known
and discussed extensively in the literature \citep[see, for
  example,][]{Tegmark1997-quadratic-estimators, Efstathiou2004-hybrid,
  CMB-mapmaking, Efstathiou2010}.  Here we provide a brief overview of the
maximum likelihood estimator~(MLE) technique and introduce our notation.
For details including discussions of invertability of the matrices, proofs
of optimality, etc., see the references.

The microwave temperature fluctuations on the sky can be represented by the
vector $\vec x(\unitvec e_j)$,
\begin{equation}
  \vec x = \mat Y\vec a+\vec n,
\end{equation}
where $\mat Y$ is the matrix of the $Y_{\ell,m}(\unitvec{e}_j)$, $j$ runs
over all pixels on the sky, $\unitvec{e}_j$ is the radial unit vector in
the direction of pixel $j$, $\vec a$ is the vector of $\alm$ coefficients,
and $\vec n$ is the noise in each pixel.  For the work considered here we
are only interested in the large-angle, low-$\ell$ behaviour so we assume
that $\vec n$ can be ignored and set $\vec n=0$ in what follows. When
working with the \WMAP{} data at low resolution this is justified, for
example the W band maps at $\nside=16$ have pixel noise
$\sigma_{\mathrm{pix}} < 3\unit{\muK}$. At higher resolution this is not as
clearly justified.  In this work we study reconstruction bias independent
of pixel noise so we may ignore $\vec n$ for our simulations.  When
setting $\vec n=0$ are further assuming that the region we are analysing is
free of foregrounds.  This is a standard, though implicit, assumption when
reconstructions are performed. The covariance matrix is then given by
\begin{equation}
  \mat C = \langle \vec x \transpose{\vec x} \rangle = \mat S.
  \label{eq:Ctheta-ensemble}
\end{equation}
Here the angle brackets, $\langle\cdot\rangle$ represent an ensemble
average.  This is the expectation value of the theoretical two point
angular correlation function, not its measured value. As is customary, we
call $\mat S$ the signal matrix despite the fact that it is \textit{not}
the two point angular correlation measured on the sky. We do not include a
noise matrix, $\mat N$, in our covariance since we are neglecting noise.

\subsection{Reconstructing the $\bmath{\alm}$}
\label{subsec:estimator-alm}

To reconstruct the $\alm$ we define the signal matrix 
as the two point
angular correlation function of \textit{the unreconstructed modes} 
\begin{equation}
  \mat C = \mat S = \sum_{\ell=\ellrecon+1}^{\ellmax} \Cl \mat P^\ell.
  \label{eq:C-alm}
\end{equation}
Here $\mat P^\ell$ is the matrix of the weighted Legendre polynomials,
\begin{equation}
  \mat P^\ell_{i,j} \equiv \frac{2\ell+1}{4\pi}
  P_\ell(\unitvec{e}_i\cdot\unitvec{e}_j),
\end{equation}
and we assume all modes with $2\le\ell\le\ellrecon$ are to be
reconstructed.  Here $\ellmax$ is the maximum multipole considered.  We
have chosen $\ellmax=4\nside+2$ for this work.  The optimal, unbiased
estimator is then given by \citep{CMB-mapmaking}
\begin{equation}
  \hat{\vec a} = \mat W\vec x, \qquad \mat W\equiv [\transpose{\mat Y}\Cinv
    \mat Y]^{-1} \transpose{\mat Y}\Cinv.
  \label{eq:estimator-alm}
\end{equation}
Note that here and throughout we work in the real spherical harmonic basis,
so $\mat Y$ is a real matrix.  The covariance matrix of our estimator is
\begin{equation}
  \Sigma \equiv \langle \hat{\vec a}\transpose{\hat{\vec a}} \rangle -
  \langle\hat{\vec a}\rangle \transpose{\langle\hat{\vec a}\rangle} 
  = [\transpose{\mat Y}\Cinv\mat Y]^{-1}.
  \label{eq:covariance-alm}
\end{equation}

The signal matrix, $\mat C$, need not include all pairs of pixels on the
sky. When it does, a reconstruction will produce precisely the spherical
harmonic decomposition.  Conversely, when a sky is masked, we only include
the unmasked pixels in $\mat C$.  The process of `masking' is thus
performed by removing the masked pixels from the signal matrix, and this
process is equivalent to assigning infinite noise to the masked pixels.

\subsection{Reconstructing the $\bmath{\Cl}$}
\label{subsec:estimator-Cl}

To reconstruct the $\Cl$ we define the signal matrix as the two point
angular correlation of \textit{all the modes};
\begin{equation}
  \mat C = \mat S = \sum_{\ell=2}^{\ellmax} \Cl \mat P^\ell.
  \label{eq:C-Cl}
\end{equation}
Notice that this differs from our previous definition (\ref{eq:C-alm}).
The optimal, unbiased estimator for
the $\Cl$ is then constructed from an unnormalized estimator, $\vec
y_\ell$.  Let
\begin{equation}
  \vec y_\ell \equiv \transpose{\vec x}\mat E^\ell\vec x, \qquad
  \mat E^\ell \equiv \frac12 \Cinv \mat P^\ell\Cinv.
  \label{eq:estimator-yl}
\end{equation}
The correlation matrix of this estimator is the Fisher matrix,
\begin{equation}
  \mat F_{\ell,\ell'} = \langle\vec y_\ell\transpose{\vec
      y_{\ell'}}\rangle - \langle\vec y_\ell\rangle \transpose{\langle \vec
      y_{\ell'}\rangle} = \frac12 \trace\bigl[\Cinv\mat P^\ell\Cinv\mat
      P^{\ell'}\bigr].
  \label{eq:Fisher-matrix}
\end{equation}
Finally, this gives the optimal, unbiased estimator of the $\Cl$,
\begin{equation}
  \hat{C}_\ell = \sum_{\ell'} \mat F^{-1}_{\ell,\ell'} \vec y_{\ell'}.
  \label{eq:estimator-Cl}
\end{equation}
Though the full Fisher matrix can be calculated, it turns out to be nearly
diagonal for reasonably small masks such as the \WMAP{} KQ75y7 mask.  In this
case the approximations
\begin{equation}
  \mat F_{\ell,\ell'} \approx \frac{2\ell+1}{2 \hat\Cl^2} \delta_{\ell,\ell'},
  \qquad \mat F_{\ell,\ell'}^{-1} \approx \frac{2 \hat\Cl^2}{2\ell+1}
  \delta_{\ell,\ell'}
  \label{eq:Fisher-approximations}
\end{equation}
may be employed.  We have confirmed the validity of this approximation and
have employed it when applicable in our subsequent analyses.

\subsection{Relating the Estimators}
\label{subsec:estimators-relation}

The optimal, unbiased estimators for $\alm$ and $\Cl$ are related to each
other.  If we define the weighted harmonic coefficients by
\begin{equation}
  \vec\beta \equiv \Sigma^{-1}\hat{\vec a},
  \label{eq:definition-beta}
\end{equation}
then 
\begin{equation}
  \vec y_\ell = \frac12\sum_m |\beta_{\ell m}|^2
  \label{eq:estimator-beta}
\end{equation}
is identical to~(\ref{eq:estimator-yl}) from which we may calculate $\hat\Cl$
\citep{CMB-mapmaking,Efstathiou2010}.

In our discussion we have been careful to note that $\mat C$ is defined
differently when used as an estimator for the $\alm$ versus the $\Cl$.  In
practise when the signal-to-noise is large the estimator for the $\alm$ is
not sensitive to the precise values and range of the $\Cl$ employed.
However, to find $\hat\Cl$ from $\hat{\vec a}$ through the weighted
harmonic coefficients~(\ref{eq:definition-beta}), the \textit{full} signal
matrix~(\ref{eq:C-Cl}) must be used when calculating the covariance
matrix~(\ref{eq:covariance-alm}) and Fisher
matrix~(\ref{eq:Fisher-matrix}).

The above discussion shows that Eq.~(\ref{eq:estimator-Cl}) is the optimal,
unbiased estimator for the $\Cl$.  Even so, given
$\hat{\vec a}$ from~(\ref{eq:estimator-alm}) it is tempting
to define a naive estimator for the $\Cl$ via
\begin{equation}
  C_\ell^e \equiv \frac1{2\ell+1}\sum_m |\hat{\vec a}_{\ell m}|^2
  \label{eq:estimator-naive-Cl}
\end{equation}
and use this to reconstruct $C(\theta)$~\cite[see fig.~5
  of][]{Efstathiou2010}.  In general this is a poor definition for the
estimator as clearly an optimal, unbiased estimator for some quantity does
\textit{not} provide an optimal, unbiased estimator for the square of that
quantity.  Its use leads to a biased estimator for the $\Cl$ and a biased
reconstruction of $C(\theta)$.  We will explore both this estimator and the
optimal, unbiased one below.

\subsection{Two Point Angular Correlation Function}

The two point angular correlation function is defined as a sky average,
that is by a sum over all pixels on the sky separated by the angle
$\cos\theta_{i,j} = \unitvec{e}_i\cdot \unitvec{e}_j$,
\begin{equation}
  C(\theta_{i,j}) \equiv \sum_{i,j} \vec x_i \vec  x_j.
\end{equation}
Ideally the two point angular correlation function would also contain an
ensemble average over realisations of the underlying model.  Since we only
have one Universe, this ensemble average cannot be calculated.  However,
for a statistically isotropic Universe the sky average and ensemble average
are equivalent.  This definition has the additional benefit that it can be
calculated on a fraction of the sky.

Alternatively the two point angular correlation function may be expanded in
a Legendre series,
\begin{equation}
  C(\theta_{i,j}) = \sum_{\ell} \frac{2\ell+1}{4\pi} C_\ell
  P_\ell(\cos\theta_{i,j}).
  \label{eq:Ctheta-vs-Cl}
\end{equation}
Note that for partial sky coverage or lack of statistical isotropy the
$C_\ell$ in this this expression are \textit{not} the same as the $\hat\Cl$
obtained from the $\alm$; see~\cite{CHSS-WMAP3} for a discussion.  This
subtlety will not be important for the following work.

\subsection{$\bmath{S_{1/2}}$ Statistic}
\label{sec:S12}

To quantify the lack of large-angle correlations the $S_{1/2}$ statistic
has been defined by \cite{WMAP1-cosmology} to be
\begin{equation}
  S_{1/2}\equiv \int_{-1/2}^1 \left[ C(\theta) \right]^2
  \dderiv(\cos\theta).
  \label{eq:S12-def}
\end{equation}
Expanding $C(\theta)$ in terms of the $\Cl$ as
above~(\ref{eq:Ctheta-vs-Cl}) we find
\begin{equation}
  S_{1/2} = \sum_{\ell,\ell'} C_\ell \mathcal{I}_{\ell,\ell'} C_{\ell'},
  \label{eq:S12-Cl}
\end{equation}
where
\begin{equation}
  \mathcal{I}_{\ell,\ell'} \equiv \frac{(2\ell+1)(2\ell'+1)}{(4\pi)^2}
  \int_{-1/2}^1 P_\ell(\cos\theta) P_{\ell'}(\cos\theta) \dderiv(\cos\theta)
\end{equation}
is a known matrix (see \citealt{CHSS-review}) that can be evaluated.

The estimator generally employed for $S_{1/2}$ is 
\begin{equation}
  \hat S_{1/2} = \sum_{\ell,\ell'} \hat C_\ell \mathcal{I}_{\ell,\ell'}
 \hat C_{\ell'}.
  \label{eq:estimator-S12}
\end{equation}
Even with $\hat \Cl$ itself   an optimal, unbiased estimator of $\Cl$, 
this does \textit{not} produce an optimal, unbiased estimator for $S_{1/2}$
\citep{Pontzen2010}.  For the unbiased estimator~(\ref{eq:estimator-Cl})
we have
\begin{eqnarray}
  \langle \hat C_\ell \hat C_{\ell'}\rangle & = &
  \sum_{\tilde\ell,\tilde\ell'} \mat F_{\ell,\tilde\ell}^{-1}
  \mat F_{\tilde\ell',\ell'}^{-1} \langle y_{\tilde\ell} \transpose{y_{\tilde\ell'}}
  \rangle \nonumber \\
  & = & 
  \sum_{\tilde\ell,\tilde\ell'} \mat F_{\ell,\tilde\ell}^{-1}
  \mat F_{\tilde\ell',\ell'}^{-1}
  \left(
  \langle y_{\tilde\ell} \rangle \transpose{\langle
    y_{\tilde\ell'}\rangle}
  + \mat F_{\tilde\ell,\tilde\ell'} \right) \nonumber \\
  & = & 
  \sum_{\tilde\ell} \left( \sum_{\tilde\ell'} \langle
  \mat F_{\ell,\tilde\ell}^{-1} y_{\tilde\ell}\rangle
  \transpose{\langle \mat F_{\ell',\tilde\ell'}^{-1} y_{\tilde\ell'}\rangle}
  + \mat F_{\ell,\tilde\ell}^{-1}\delta_{\ell',\tilde\ell}
  \right)\nonumber \\
  & \approx & C_\ell C_{\ell'} + \frac{2}{2\ell+1} C_\ell^2 \delta_{\ell,\ell'}.
\end{eqnarray}
In the second line we have used the definition of the Fisher
matrix~(\ref{eq:Fisher-matrix}), the third line is an algebraic
simplification, and in the final line we have again
used~(\ref{eq:estimator-Cl}), the fact that $\hat C_\ell$ is
unbiased, and the approximation from~(\ref{eq:Fisher-approximations}).
With this it now straightforward to see that
\begin{eqnarray}
  \langle\hat S_{1/2}\rangle & = &
  \sum_{\ell,\ell'} \langle \hat C_\ell \hat C_{\ell'}\rangle
  \mathcal{I}_{\ell,\ell}
  = S_{1/2} + \sum_{\ell} \frac{2 C_\ell^2}{2\ell+1}
  \mathcal{I}_{\ell,\ell}
  \nonumber \\
  & \neq & S_{1/2}.
\end{eqnarray}
It is thus clear that~(\ref{eq:estimator-S12}) is a biased estimator and, in
fact, is biased toward larger values of $S_{1/2}$.

As noted by \cite{Pontzen2010} this is of `pedagogical interest' but does
not affect the studies of low $S_{1/2}$.  The Monte Carlo simulations
employed \citep[see][for example]{CHSS-WMAP5} account for this bias.  It
does suggest that an alternative measure of the lack of large-angle
correlations is desirable.

\subsection{Assumptions}

\cite{Efstathiou2010} claim that the full-sky, large-angle CMB can be
reconstructed solely from the harmonic structure of the CMB outside the
masked, Galactic region, and independent of the contents of the
masked portion of the sky. We will demonstrate in what follows that this
claim does not hold up to closer scrutiny.

It is clear that without assumptions regarding the harmonic structure inside
the masked region nothing can be said about it.  In principle the low-$\ell$
harmonic structure inside the masked region could be anything, ranging from no
power, to large power, to wild oscillations, making the full-sky
reconstruction impossible.  

Assuming a cosmological origin for the observed microwave signal outside
the masked region, it seems reasonable to assume it will be consistent with
the signal inside the masked region.  With that assumption, the harmonic
structure outside the masked region can be extended into the masked region.
For actual, full-sky maps there is a further assumption: the region inside
the mask is well enough determined and statistically close enough to the
region outside the mask that it does not bias the reconstruction.  This
latter assumption turns out to not be true as we demonstrate below.

Note also that if the region inside the mask is trusted, then there is no
need to perform either masking or the reconstruction at all, the full-sky
map can be analysed directly. Therefore, validity of the stringent
assumptions required for the reconstruction obviates the very need for the
reconstruction.

When the reconstruction formalism described above is applied to actual
data, further assumptions are implicit.  In our development we have
assumed that the temperature fluctuations contain pure CMB signal. In
practise, besides pixel noise (which we have not included as described
above) the data may contain unknown foregrounds.  To avoid contamination
by foregrounds it is common to analyse a foreground-cleaned map, such as
the ILC map, and to mask the most contaminated regions of the sky.
In following this approach, care must be taken not to reintroduce
contamination in the data prior to reconstruction.  As we will show below,
the standard process of preparing data for reconstruction, in particular
smoothing the full-sky map, violates this requirement.

\section{Results}
\label{sec:results}

To explore how data handling prior to reconstruction affects the results,
we have performed a series of Monte Carlo simulations of $\Lambda$CDM based
on reconstruction procedures suggested in the literature.  We have employed
the simplest best-fitting $\Lambda$CDM model from \WMAP{} based solely on
the \WMAP{} data.  This is model ``lcdm+sz+lens'' with ``wmap7'' data from
the lambda site.  Our results are insensitive to the exact details of the
model since we are performing a theoretical study examine relative
differences between reconstructions and not performing parameter
estimation.  Our simulations are performed at $\nside=128$ unless otherwise
noted and we will focus on the reconstruction of $a_{2 m}$ and $C_2$.
Further, our simulations only consider $\ellrecon=10$, reconstruct from the
pixels outside the KQ75y7 mask provided by \WMAP{} and degraded to the
appropriate resolution, and use the data from the \WMAP{} seven year
release.

A collection of realisations of the full sky are created as follows:
\begin{enumerate}
\item Generate a random sky at $\nside=512$ from the best-fitting
  $\Lambda$CDM model.
\item Extract the $a_{2 m}$ and calculate the power in the quadrupole,
  denote this value by $C_2$.
\item Rescale the $a_{2 m}$ so that the $C_2$ in the map has a fixed value,
  for example, rescale so that $C_2=100\unit{(\muK)^2}$ by replacing the
  $a_{2 m}$ with $a_{2 m} \rightarrow a_{2 m}
  \sqrt{100\unit{(\muK)^2}/C_2}$.  Notice that this does not change
  the phase structure of the $a_{2 m}$.
\item Smooth the map with a $10\degr$ Gaussian beam, if desired.
\item Degrade the map to the desired resolution ($\nside=128$ or $\nside=16$). 
\item Repeat the rescaling of the $a_{2 m}$ for each value of $C_2$
  that we wish to consider. In our simulations we consider
  $C_2=10\mbox{--}10^4\unit{(\muK)^2}$.
This ensures that the same map realisation is used with only the quadrupole
power changed. 
\end{enumerate}
This procedure constitutes a single realisation. The results in this work
are based on at least $20,000$ realisations.

Degrading masks requires an extra processing step.  Pixels near mask
boundaries turn from the usual $1$ or $0$ to denote inclusion or exclusion
from the analysis, respectively, to fractional values.  We redefine our
degraded masks by setting all pixels with a value greater than $0.7$ to $1$
and all others to $0$. For the KQ75y7 mask this process leaves about $70$
per cent of the pixels for analysis.  To be precise, at $\nside=128$ this
leaves $136,828$ unmasked pixels and at $\nside=16$ there are $2,157$
pixels left.

A map with a modest angular resolution contains all the low-$\ell$ CMB
information, so it may seem surprising that we employ $\nside=128$ in our
studies.  Instead it is common in low-$\ell$ studies to employ a map at
$\nside=16$, corresponding to pixels of approximately $3\degr$ in size
\citep[see][for a recent example of this]{Efstathiou2010}.  The effects of
the choice of resolution, the need for smoothing a map prior to analysis, and
the leaking of information this causes will now be explored.

\subsection{Choice of Map Resolution}

\begin{figure}
  \includegraphics[width=0.5\textwidth]{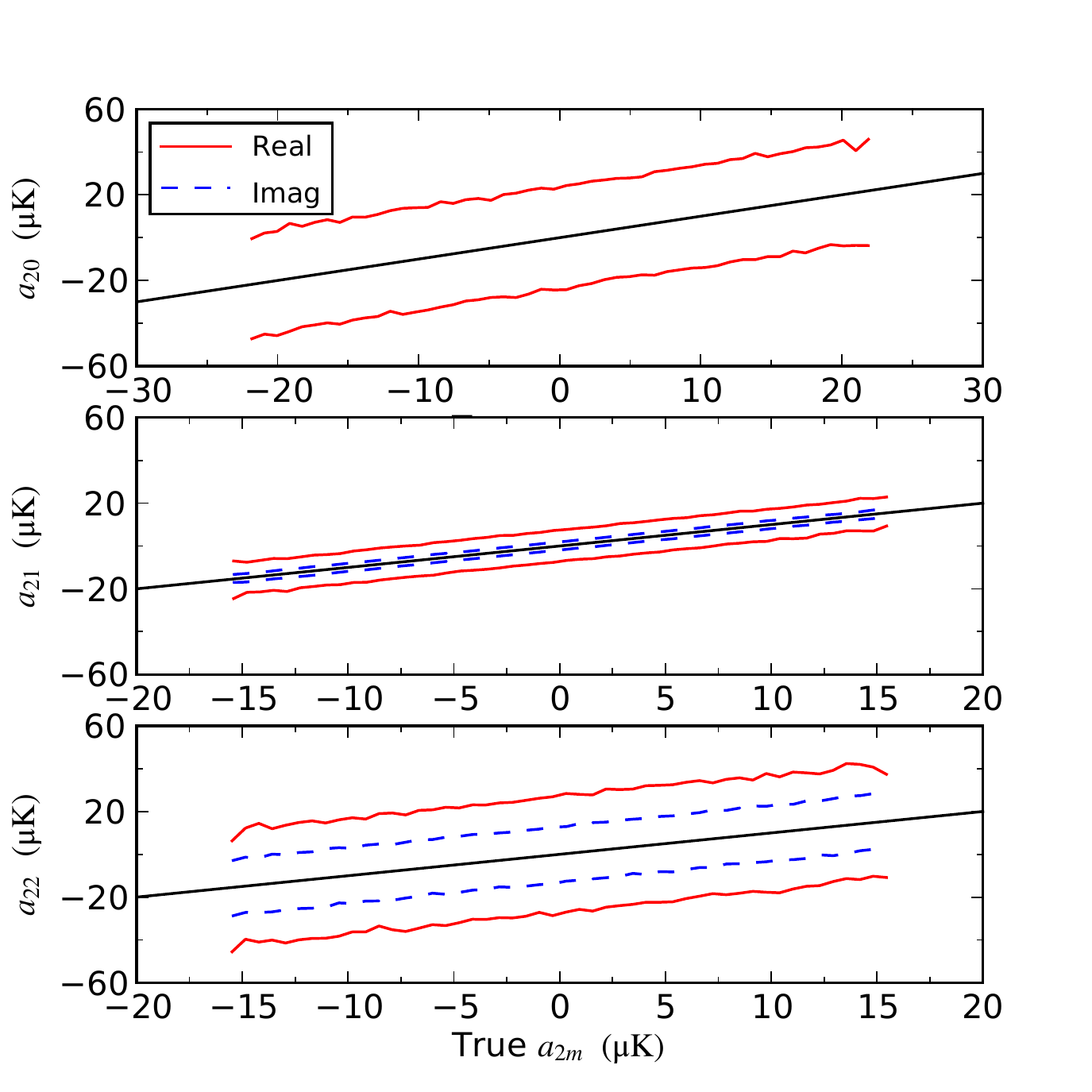}
  \caption{The $95$ and $5$ percentile lines for the $a_{2 m}$
    reconstructed from the pixels outside the KQ75y7 mask at $\nside=128$ (and
    thus $\ellmax=514$) of $\Lambda$CDM realisations with
    $C_2=100\unit{(\muK)^2}$, as discussed in the text.  The red, solid
    lines are for the real part of the $a_{2 m}$ and the blue, dashed lines
    are for the imaginary part.  The black, solid line shows the expected
    result for a perfect reconstruction.  We see that the reconstruction is
    unbiased, that is, it tracks the true value.}
  \label{fig:recon-a2m-C2-100-nside-128}
\end{figure}

\begin{figure}
  \includegraphics[width=0.5\textwidth]{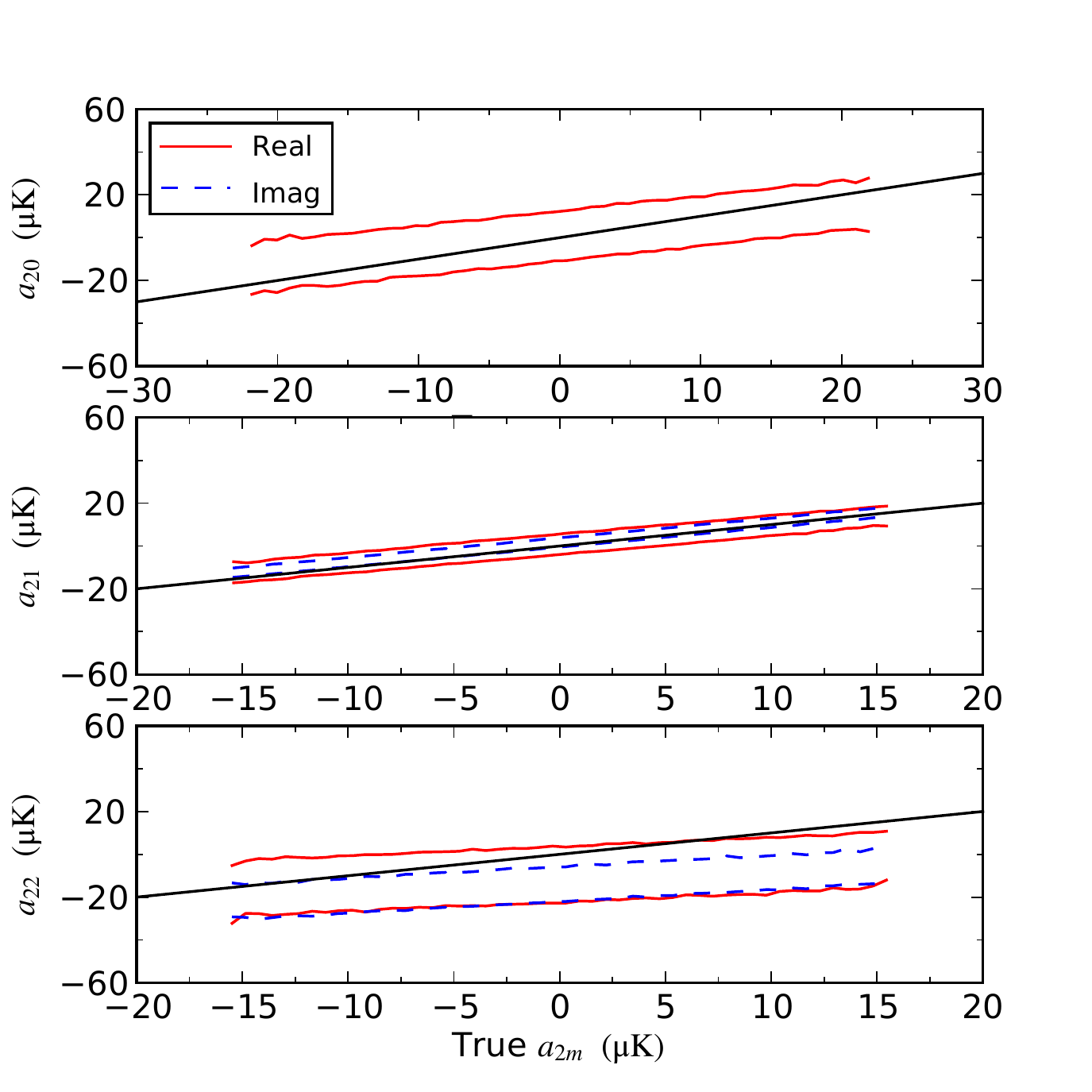}
  \caption{The same as Fig.~\ref{fig:recon-a2m-C2-100-nside-128} now
    reconstructed from the pixels outside the KQ75y7 mask at $\nside=16$
    (and thus $\ellmax=66$).  Here we see that the reconstruction is
    biased.}
  \label{fig:recon-a2m-C2-100-nside-16}
\end{figure}

The study of large-angle, low-$\ell$ properties of the CMB appears naively
not to require high resolution maps. Maps degraded to the resolution
corresponding to $\nside=16$ are commonly employed
\citep{CMB-mapmaking,Efstathiou2010}.  When a map is degraded by averaging
over pixels, high frequency noise is introduced as may be seen in
Figs.~\ref{fig:recon-a2m-C2-100-nside-128} and
\ref{fig:recon-a2m-C2-100-nside-16}.  These figures show the reconstructed
$a_{2 m}$ using the optimal, unbiased estimator from
Eq.~(\ref{eq:estimator-alm}) for realisations with
$C_2=100\;\unit{(\muK)^2}$. The solid, red lines (dashed, blue lines) show
the $5$ and $95$ percentile lines from our realisations for the
reconstructed real (imaginary) parts of each $a_{2 m}$, using maps degraded
to $\nside=128$ (Fig.~\ref{fig:recon-a2m-C2-100-nside-128}) and $\nside=16$
(Fig.~\ref{fig:recon-a2m-C2-100-nside-16}) and pixels outside the KQ75y7
mask.  As expected from an unbiased estimator the reconstructed values
track the true values (Fig.~\ref{fig:recon-a2m-C2-100-nside-128}).  Further
we see that the $a_{2 1}$ are best determined and the $a_{2 0}$ and
$a_{2 2}$ have larger variances due to the mask which produces greater
admixture of ambiguous modes for these cases.  However, for $\nside=16$
(Fig.~\ref{fig:recon-a2m-C2-100-nside-16}) we see that the reconstruction
does not track the true values and is instead biased.  This bias is due to
the averaging done to degrade the maps and becomes more significant the
more the map is degraded. From this we conclude that the coupling of the
small-scale modes to the large-scale modes caused by using maps with
resolution that is too coarse can be at least partly responsible for
reconstruction bias.

\subsection{Smoothing the Map}

\begin{figure}
  \includegraphics[width=0.5\textwidth]{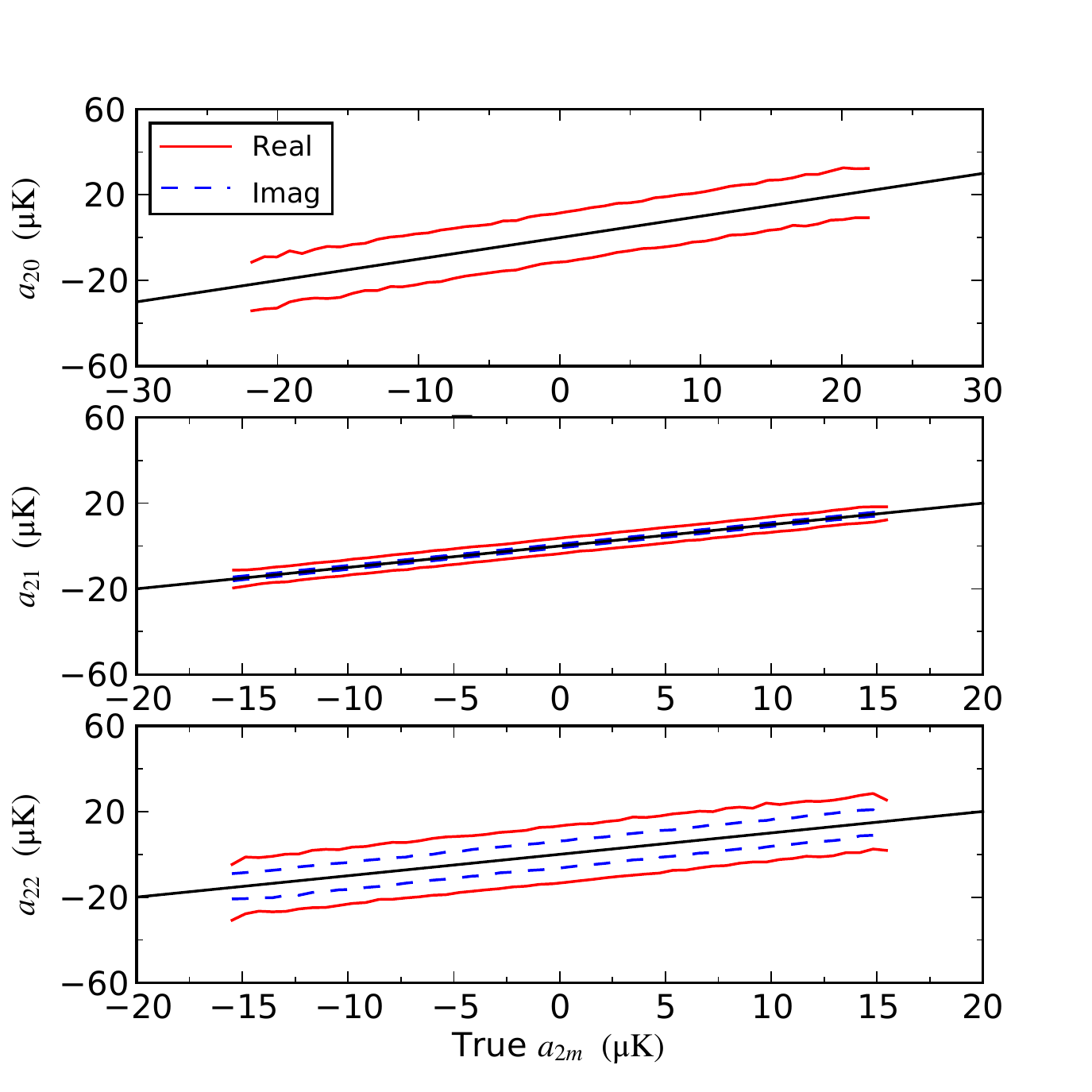}
  \caption{The same as Fig.~\ref{fig:recon-a2m-C2-100-nside-128} now
    reconstructed from maps smoothed with a $10\degr$ Gaussian beam
    applied to the full sky map.  As in
    Fig.~\ref{fig:recon-a2m-C2-100-nside-128}, the reconstruction is
    unbiased.}
  \label{fig:recon-a2m-C2-100-nside-128-10deg}
\end{figure}

\begin{figure}
  \includegraphics[width=0.5\textwidth]{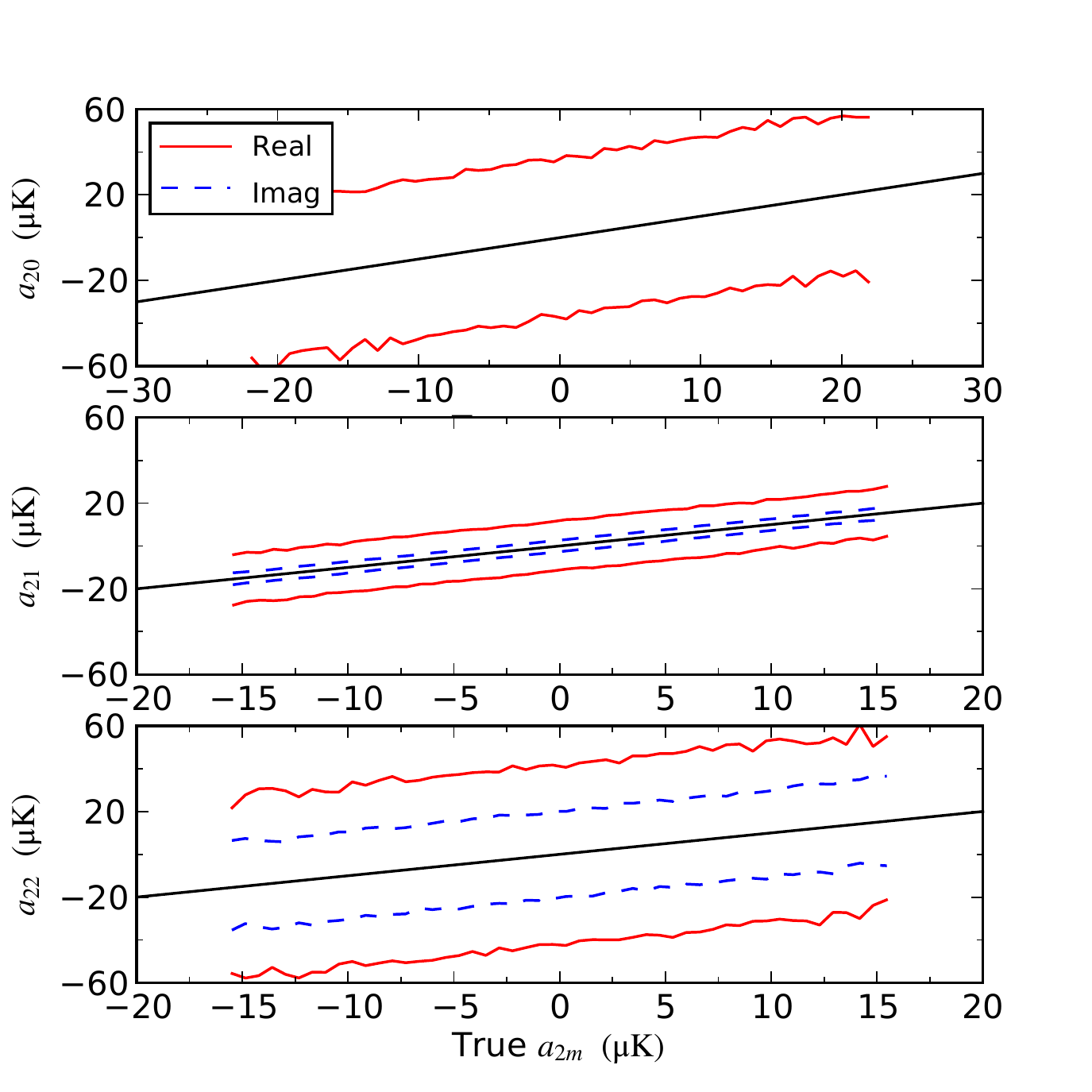}
  \caption{The same as Fig.~\ref{fig:recon-a2m-C2-100-nside-128-10deg} now
    reconstructed from maps with $\nside=16$. The reconstruction is now
    also unbiased.}
  \label{fig:recon-a2m-C2-100-nside-16-10deg}
\end{figure}

In practise raw degraded maps are not used for the reasons shown in the
previous section, instead the maps are smoothed with a Gaussian beam with
FWHM of at least the size of the pixels and then degraded.  In this work we
employ a smoothing scale of $10\degr$, consistent with
\cite{Efstathiou2010}.  Smoothing the maps studied in the previous section
prior to reconstructing the $\alm$ produces the results shown in
Figs.~\ref{fig:recon-a2m-C2-100-nside-128-10deg} and
\ref{fig:recon-a2m-C2-100-nside-16-10deg}.  With smoothing we see that the
$\alm$ estimator is unbiased for both resolutions, $\nside=128$ and
$\nside=16$.  Smoothing is thus an essential step when working with low
resolution maps.

In Figs.~\ref{fig:recon-a2m-C2-100-nside-128-10deg} and
\ref{fig:recon-a2m-C2-100-nside-16-10deg} we also see that the variance in
the reconstructed values is resolution dependent with the smaller variance
provided by the higher resolution maps.  Again this is not surprising, and
can be understood as follows.  Our covariance matrix in
Eq.~(\ref{eq:covariance-alm}) does not include a noise term yet we have
introduced noise by degrading.  Smoothing does a good job at reducing the
noise to a level where the reconstruction is unbiased, however, there is
still residual noise that affects the covariance of the estimator.  The
higher the resolution the smaller this noise.  The best results are
obtained by working at the highest resolution that is feasible.  For this
reason we work at $\nside=128$ in our simulations.  See
Appendix~\ref{app:high-resolution} for technical details.

\subsection{Reconstructing the $\bmath{\alm}$}

We have now seen that the estimator in Eq.~(\ref{eq:estimator-alm}) is an 
optimal, unbiased estimator for the $\alm$ when we work at high resolution 
and/or smooth the maps prior to reconstruction
(Figs.~\ref{fig:recon-a2m-C2-100-nside-128},
\ref{fig:recon-a2m-C2-100-nside-128-10deg},
\ref{fig:recon-a2m-C2-100-nside-16-10deg}). Although
this has only been shown for $C_2=100\unit{(\muK)^2}$ we have verified that
this is true independent of the quadrupole power.

\begin{figure}
  \includegraphics[width=0.5\textwidth]{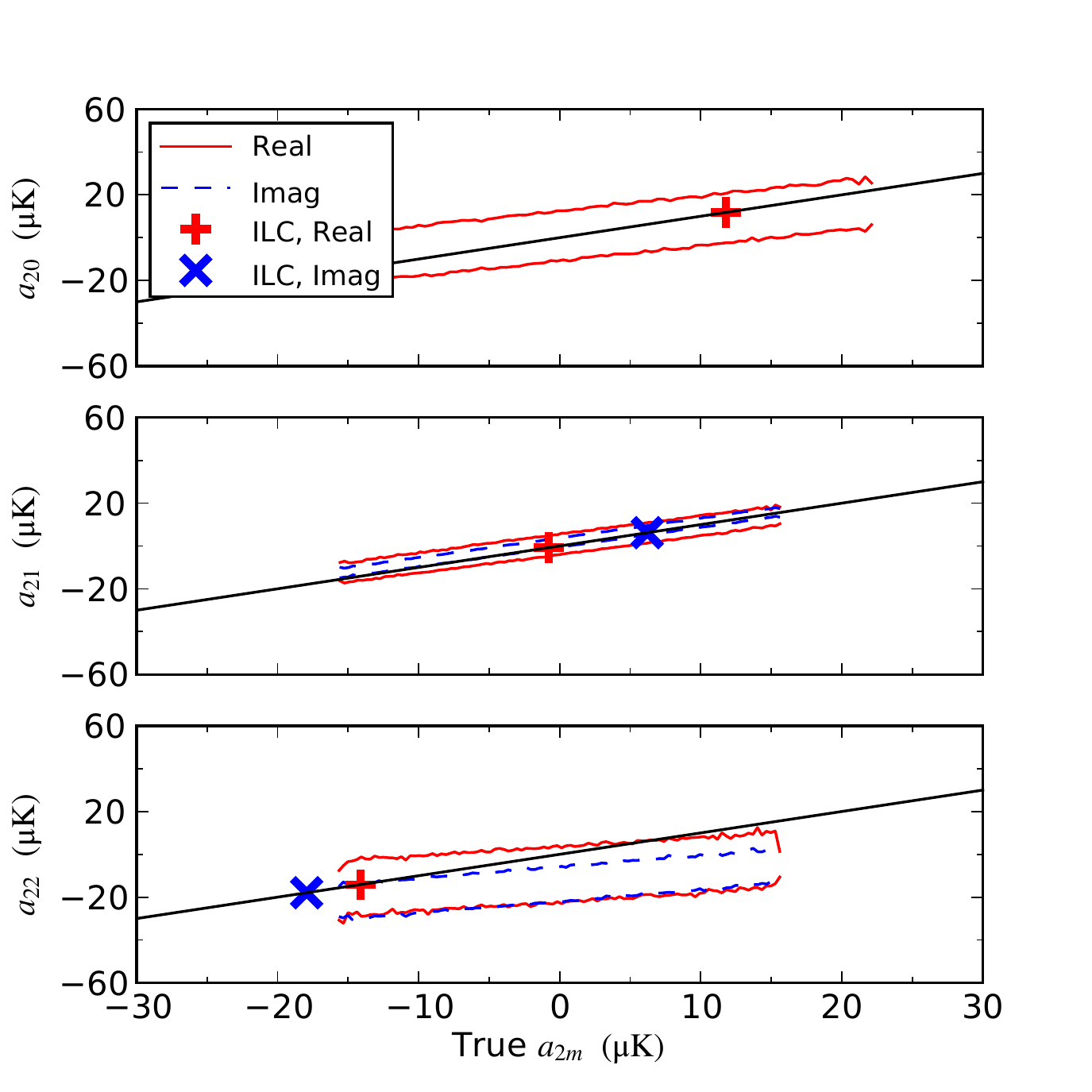}
  \caption{The same as Fig.~\ref{fig:recon-a2m-C2-100-nside-128-10deg} now
    with the masked region filled in with the ILC map \textit{prior} to
    smoothing and rescaling.  We clearly see the reconstructed $a_{2 m}$
    are \textit{not} unbiased.  The bias in reconstructing $a_{20}$ and
    $a_{22}$ is particularly apparent.  This is due to the leakage of
    information from inside the masked region.}
  \label{fig:recon-a2m-C2-100-with-ILC}
\end{figure}

\begin{figure}
  \includegraphics[width=0.5\textwidth]{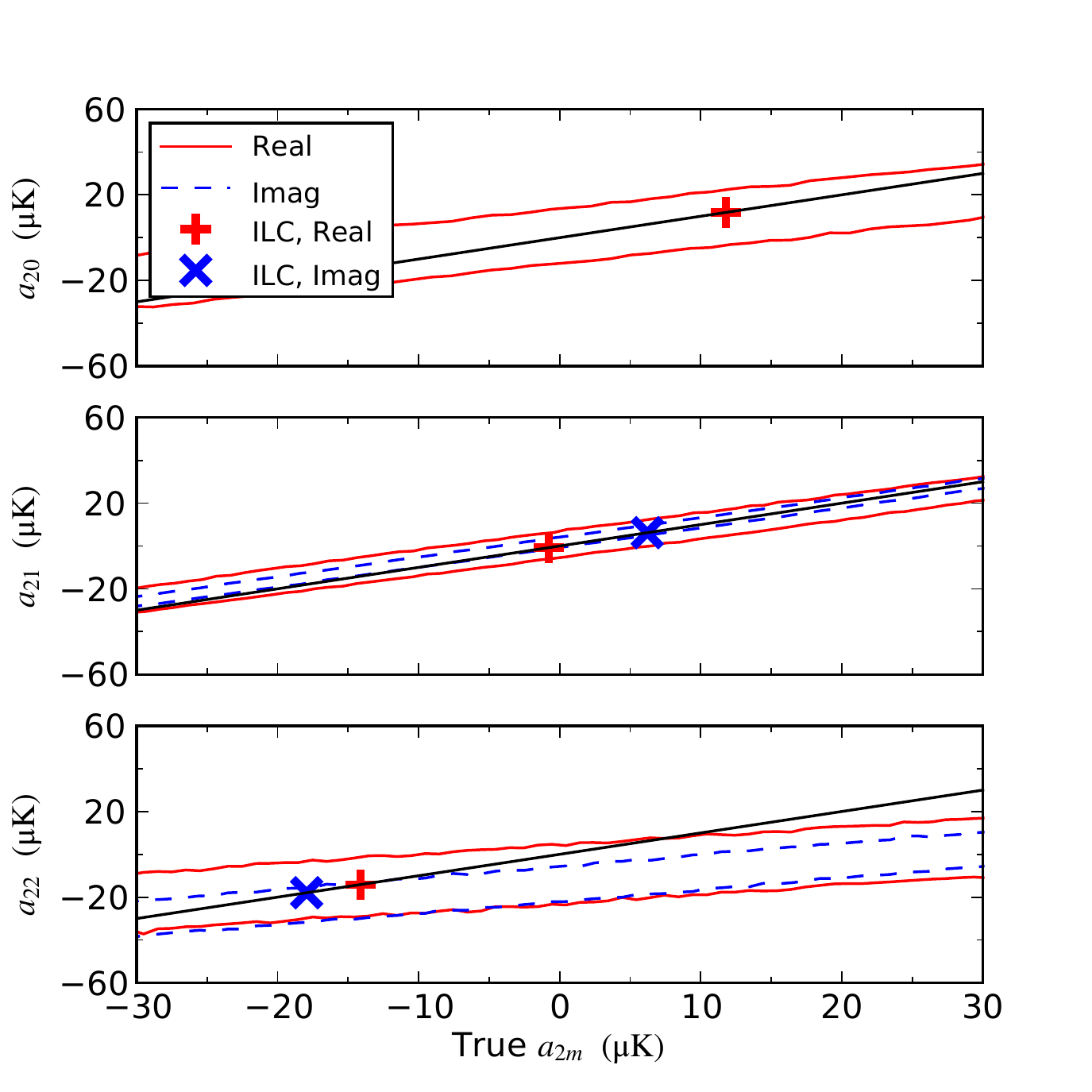}
  \caption{The same as Fig.~\ref{fig:recon-a2m-C2-100-with-ILC} now with
    $C_2=1000\unit{(\muK)^2}$.}
  \label{fig:recon-a2m-C2-1000-with-ILC}
\end{figure}

As noted above, the fact that we are smoothing the maps prior to masking
imposes assumptions on the maps.  For the realisations discussed above the
assumptions are met; the region inside the mask is, statistically, identical
to the region outside.  However, for real data the Galaxy is a bright
foreground that must be removed.  The \WMAP{} ILC procedure attempts to do this
and produce a full-sky CMB map.  Even so, masking is often performed to avoid
relying on the information inside this region since it may still be
contaminated by Galactic foregrounds.  

Unfortunately, when the map is smoothed information leaks out of the masked
region and biases the reconstruction as shown in
Figs.~\ref{fig:recon-a2m-C2-100-with-ILC} and
\ref{fig:recon-a2m-C2-1000-with-ILC}.  For this analysis, for each synthetic
map we filled the 
masked region with the corresponding portion (i.e.\ the masked region)
taken from the ILC map.  We then smoothed and degraded the resulting
synthetic map. In these two figures we show the true and reconstructed
values of the coefficients $a_{2 m}$; we also show the ILC map's $a_{2
  m}$ for reference.  We clearly see the bias in the reconstructed
$a_{2 m}$ and its correlation with the ILC values.  If $a_{2
  m}^{\mathrm{rec}}<a_{2 m}^{\mathrm{ILC}}$, then $a_{2
  m}^{\mathrm{rec}}$ is biased upwards, and vice versa. For example,
the ILC $a_{2 2}$ values are large and negative which leads to the
reconstruction being skewed to agree better at large, negative values
than at large positive values.  This trend continues for the other
$a_{2 m}$ and clearly shows that the smoothing has mixed information
from the masked region.  

We can also recognise other details in the quality of the reconstruction
that are specifically due to the orientation of the KQ75y7 mask in Galactic
coordinates. For example, we see that the variance in the reconstructed
real part of $a_{22}$ is larger than that for the imaginary part
of $a_{22}$; the reason is that the real part of $Y_{22}$ has an extremum in
the centre of the Milky Way where the mask `bulges' while the imaginary
part has a node at this location. Therefore, more information relevant to
the real part of $a_{22}$ is missing than for the imaginary part, and the
former has a larger reconstruction error. Moreover, it is also the case
that the $Y_{20}$ and $Y_{22}$ have extrema in the Galactic plane whereas
$Y_{21}$ has nodes.  Due to this the variances of $a_{20}$ and $a_{22}$ are
expected to be larger than that of $a_{21}$, as our reconstruction plots
show.

Notice also that the reconstruction bias we find is \textit{not} an artifact
of the sharp transition introduced in the process of filling the masked
regions of simulated maps with the ILC contents. The smoothing procedure,
for one, completely removes the sharp feature in the map. Moreover, we have
explicitly checked that the reconstructed $\alm$ are not biased when the
cut is filled with contents of \textit{another} statistically isotropic
map. Therefore, the reconstruction bias seen in our plots is real, and is
caused to the specific structure of the ILC map behind the Galactic plane
which `leaks' into the unmasked region.

The question, then, is how to fill the masked region before smoothing.
In principle anything could be used to fill the Galactic region, but
then the information about this fill would leak outside the mask due
to the smoothing.  If the map were masked prior to smoothing then
`zero' would be leaked and bias the reconstruction.  Alternatively, if
the Galactic region were filled with Gaussian noise with
root-mean-squared value consistent with the region outside the Galaxy
then the estimator would be unbiased similar to the results in
Fig.~\ref{fig:recon-a2m-C2-100-nside-128}, but this would rely on the
\textit{assumption} that the true CMB inside the mask has precisely
the same statistical properties as the CMB in the region outside.
Filling with the ILC values would make sense if we could be completely
confident that the ILC reconstruction of the region inside the mask is
accurate.  However, in the ILC the region inside the Galactic mask has
different statistical properties than the region outside, particularly
for the large-angle behaviour.  This alone raises concerns that the
ILC reconstruction is not entirely accurate.  Further, if we knew how
to properly treat the region inside the mask, either by accepting the
ILC values or filling it with appropriate statistical values, there
would be no need for a reconstruction as we would have a full sky map
to analyse!

The challenge is that there is no, or at least no unique, compelling choice
of how to fill the masked region before smoothing.  In the face of this,
the approach we take below is to study how the admixture of the large-angle
behaviour of the Galactic region from the ILC map affects the
reconstruction of the low-$\ell$ CMB, particularly when the region outside
the Galaxy has low quadrupole power and lack of large-angle correlations.
We show how this particular choice biases the reconstruction.

\subsection{Reconstructing the $\bmath{\Cl}$}

Since we are interested in reconstructing $C(\theta)$ we next need to
reconstruct the $\Cl$.  From the $\alm$ we first proceed using the naive
estimator~(\ref{eq:estimator-naive-Cl}), denoted $\Cl^e$ (as used to
generate fig.~5 of \citealt{Efstathiou2010}).

\begin{figure}
  \includegraphics[width=0.5\textwidth]{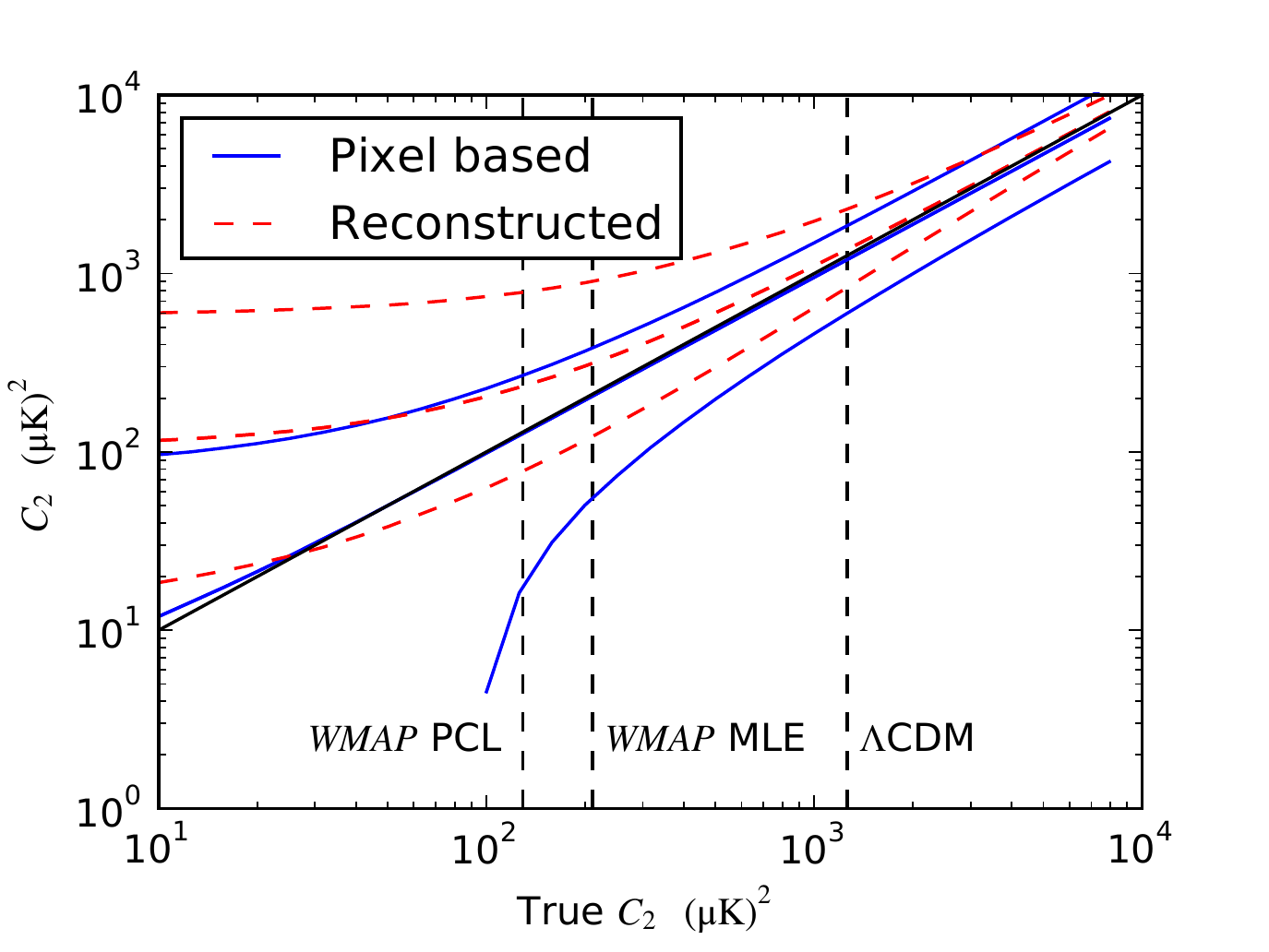}
  \caption{The $95$, $50$, and $5$ percentile lines of the reconstructed
    $C_2$ (top to bottom, respectively) from our realisations. The maps
    have \textit{not} been smoothed prior to reconstruction. The pixel based
    (blue, solid line) comes from \spice; where as, the reconstructed (red,
    dashed line) is the estimator $\Cl^e$.  We see that this estimator is
    clearly biased toward larger reconstructed values for small, true
    $C_2$, such as the values extracted from \WMAP{} using either the PCL
    or MLE procedures.  For a value of $C_2$ near the $\Lambda$CDM value
    the reconstruction method is a good estimator. The pixel based method
    produces values of $C_2$ with a median much closer to the true values,
    though with larger error bars.}
  \label{fig:recon-C2-r7}
\end{figure}

The results for this estimator are shown in Fig.~\ref{fig:recon-C2-r7}.
For these realisations the maps were \textit{not} smoothed.  The
reconstruction is shown as the dashed, red lines representing the $5$,
$50$, and $95$ percentile values as a function of the true $C_2$ used to
generate the maps.  The solid, blue lines are the equivalent values from
the reconstruction based on the pixel estimator from \spice.\@ Again the
solid, black line is the reconstructed=true relation plotted to guide the
eye.  At large $C_2$ we see the desired behaviour: the reconstructed values
from both estimators are centred around the true value, and $C_2^e$ does
have a smaller variance, as an optimal estimator should (however, this does
not mean it \textit{is} optimal).  At low $C_2$, in particular near the
\WMAP{} PCL and MLE values, the pixel based estimator is
still centred around the true value, though with large variance; however,
the $C_2^e$ is now biased toward larger values.

\begin{figure}
  \includegraphics[width=0.5\textwidth]{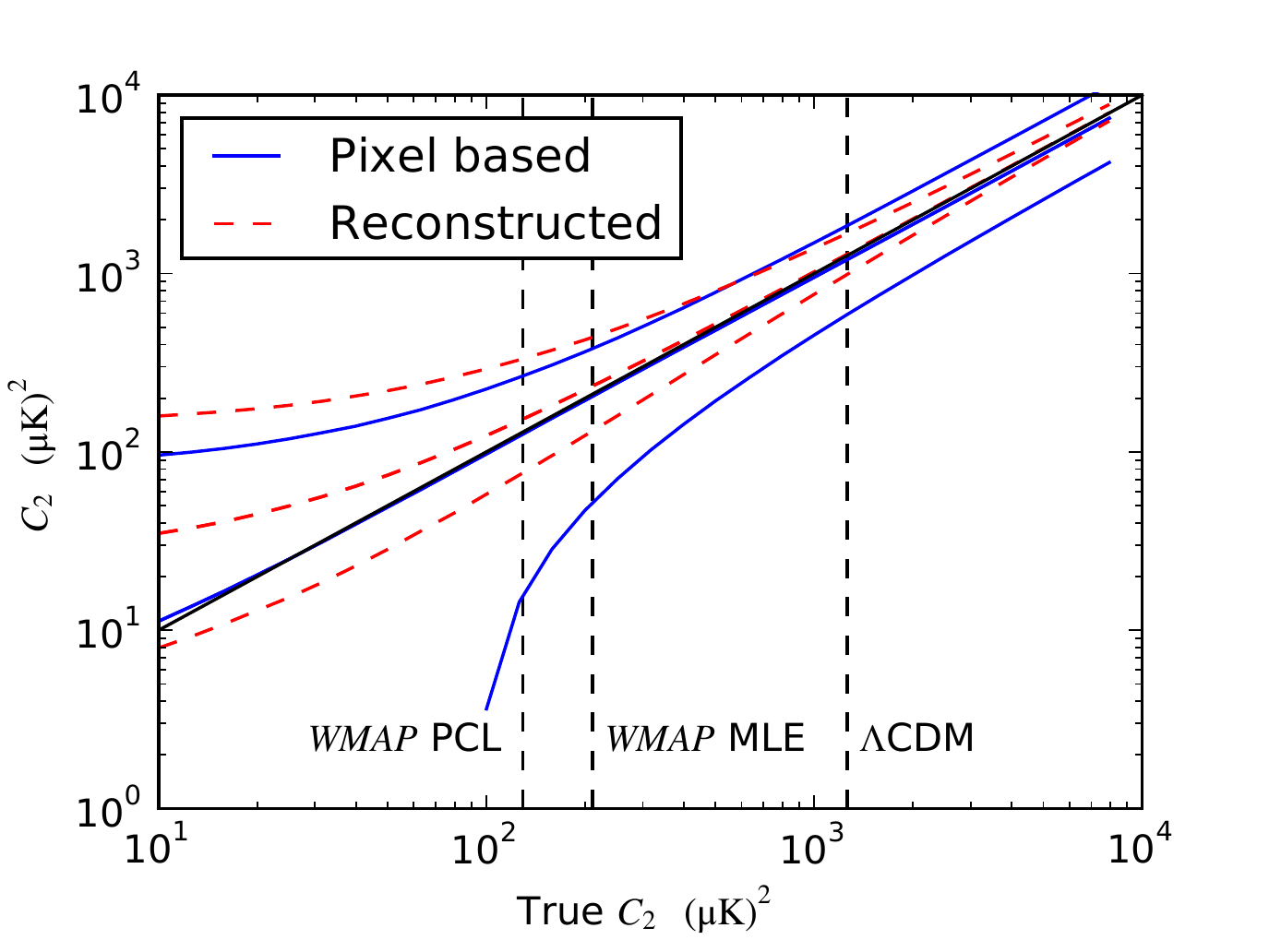}
  \caption{The same as Fig.~\ref{fig:recon-C2-r7} now with the realisations
    smoothed to $10\degr$ prior to reconstruction.  It appears that the
    estimator~(\ref{eq:estimator-naive-Cl}) does a better job of
    reproducing $C_2$ for a $\Lambda$CDM model, though, see
    Fig.~\ref{fig:recon-C2-r7-10deg-ILC}.}
  \label{fig:recon-C2-r7-10deg}
\end{figure}

The results in Fig.~\ref{fig:recon-C2-r7} were for unsmoothed maps.  The
usual approach is to smooth the maps which suppresses power on small
scales (high-$\ell$).  Fig.~\ref{fig:recon-C2-r7-10deg} shows the results
when the maps are smoothed prior to reconstruction; they are encouraging.
Both estimators now track the true values much more closely.  Even the
median of $C_2^e$ remains close to the true value for values near the \WMAP{}
PCL value.  This shows that with smoothing the correlations are reduced due
to the lack of high frequency noise.  It suggests that smoothing the map,
reconstructing the $\alm$, and employing $\Cl^e$ as our estimator is
sufficient and nearly optimal.

\begin{figure}
  \includegraphics[width=0.5\textwidth]{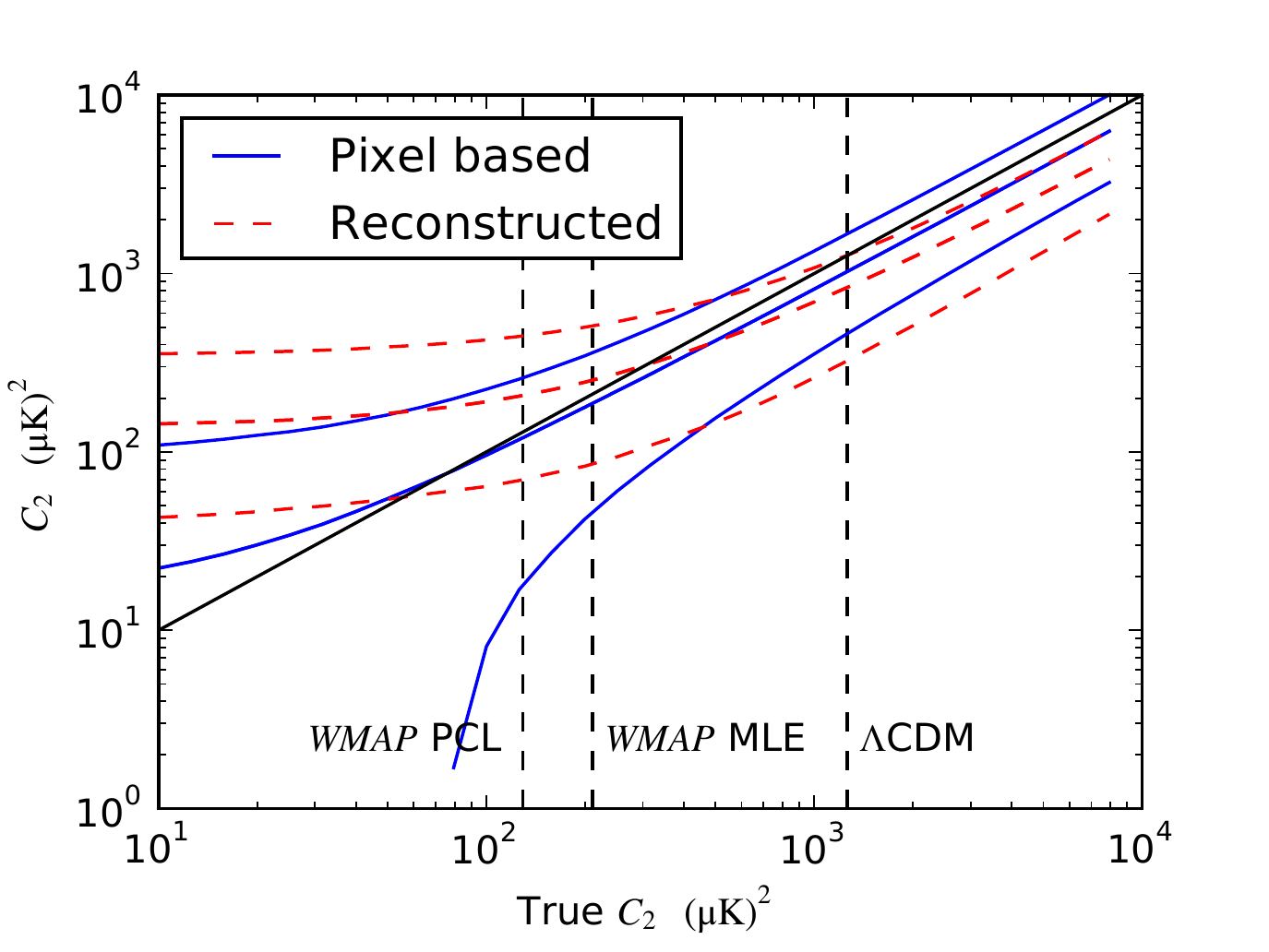}
  \caption{The same as Fig.~\ref{fig:recon-C2-r7-10deg} now with the region
    inside the masked replaced by the ILC prior to smoothing to $10\degr$
    and reconstructing.  Clearly the ILC information from inside the masked
    region has leaked out biasing the reconstruction.  Not surprisingly the
    reconstruction now is only accurate near the \WMAP{} MLE value; the value
    consistent with this region of the ILC. At lower $C_2$ values the
    reconstruction plateaus to this value as it is the main contribution to
    quadrupole power. At higher values of $C_2$ the quadrupole power is
    suppressed by the leakage.}
  \label{fig:recon-C2-r7-10deg-ILC}
\end{figure}

Unfortunately this is not the case.  As noted above, smoothing makes
assumptions about the validity of the region inside the mask.  We saw that
even for the $\alm$ this leads to a bias (see
Fig.~\ref{fig:recon-a2m-C2-100-with-ILC}). When the corresponding ILC
portion is placed into the masked region prior to smoothing the $C_2^e$ is
also biased as shown in Fig.~\ref{fig:recon-C2-r7-10deg-ILC}.  We see that
the masked region drives $C_2^e$ to be near the value inside the mask
(approximately the \WMAP{} MLE value).  The $C_2^e$ results are biased
upward for very small $C_2$ and downward for large $C_2$.  Thus, even
though smoothing helps in removing the correlation bias in the $\Cl^e$
estimator it introduces its own bias.  How the masked region is filled
determines how the distribution of $C_2^e-C_2^{\mathrm{true}}$ will be
skewed.  Roughly speaking the values inside the mask will be favoured,
raising the reconstructed values that are lower than the masked region
values, and lowering values that are higher than those from the masked
region.

We have seen that the naive estimator, $\Cl^e$, provides an unbiased
estimate of $C_2$ when the true value is near the expected, $\Lambda$CDM
value.  However, when the true value is low this estimator tends to
\textit{overestimate} $C_2$.  Further, when smoothing is applied the
reconstruction skews the values towards those consistent with the region
inside the mask.  This is to be expected.  In fact, if the region inside
the mask were believed then there would be no need to reconstruct at all, a
full-sky map would already exist and it could be used for analysis without
this extra effort.

\begin{figure}
  \includegraphics[width=0.5\textwidth]{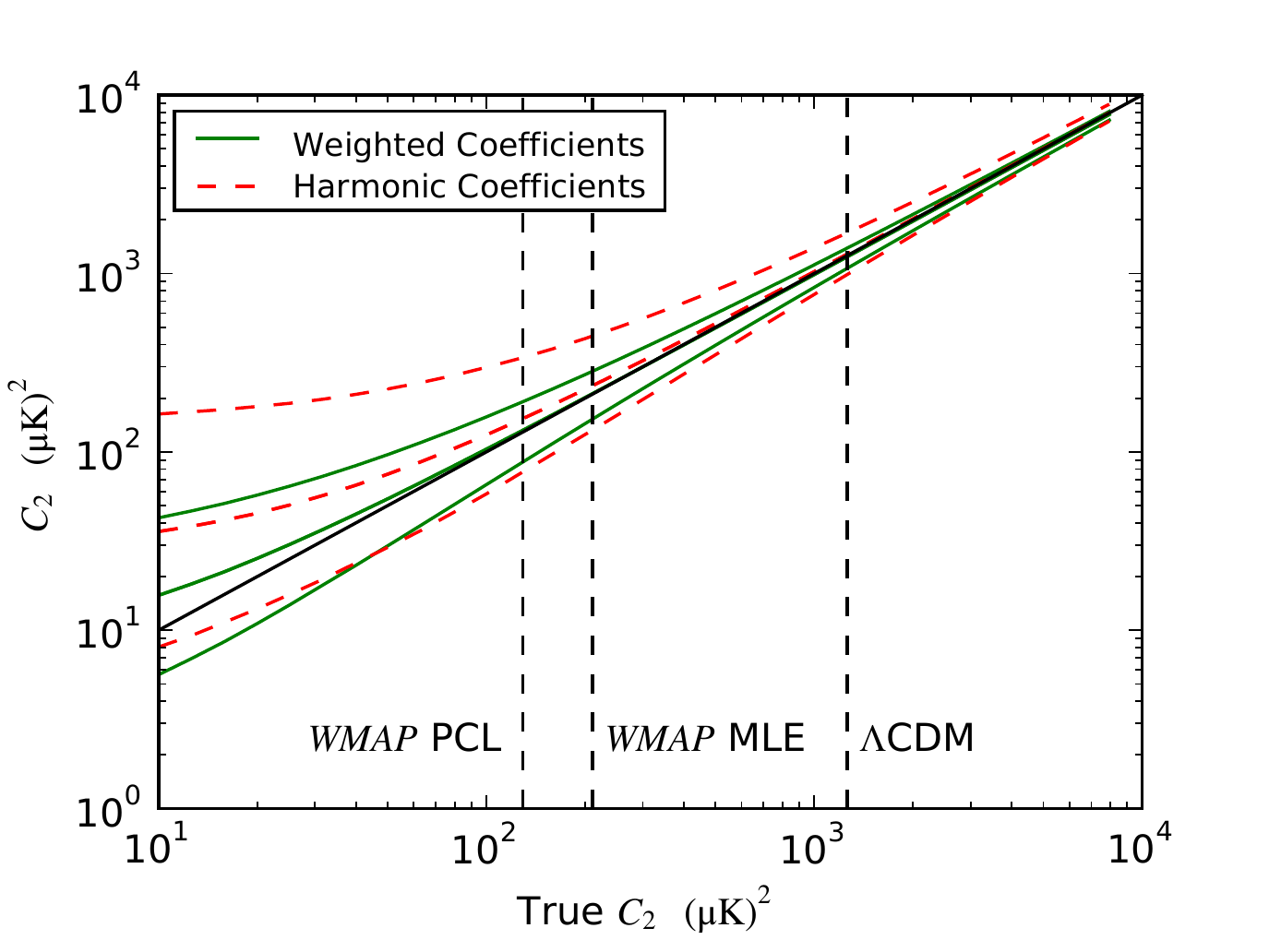}
  \caption{Similar to Fig.~\ref{fig:recon-C2-r7-10deg} now comparing
    $C_2^e$, the harmonic coefficient
    estimator~(\ref{eq:estimator-naive-Cl}) again as the dashed, red
    lines to $\hat C_2$, the weighted harmonic coefficient
    estimator~(\ref{eq:estimator-Cl}) as the green, solid lines.  We see
    that the weighted harmonic coefficient estimator is unbiased over the
    full true $C_2$ range.}
  \label{fig:recon-unbiased-C2-r7}
\end{figure}

\begin{figure}
  \includegraphics[width=0.5\textwidth]{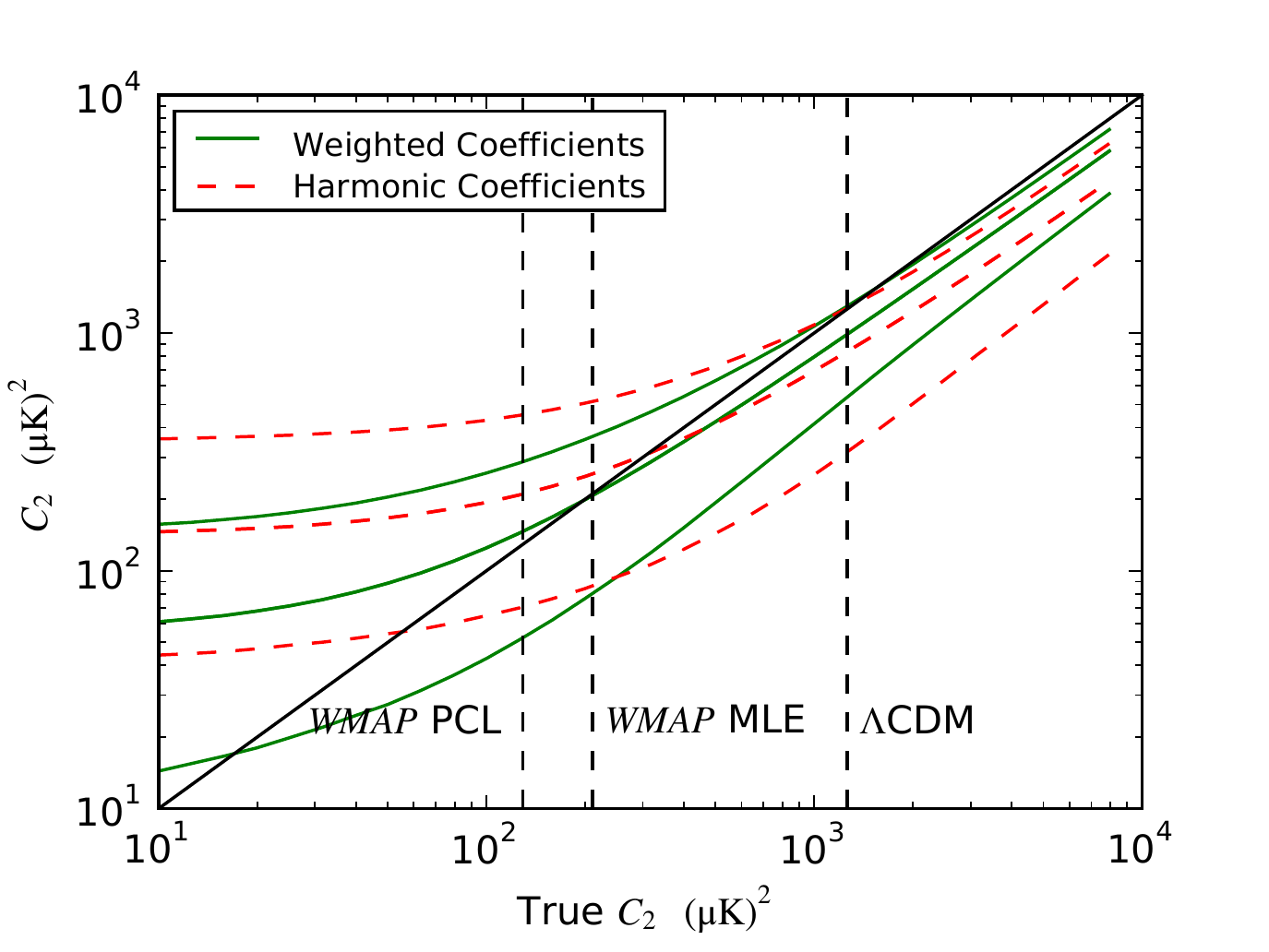}
  \caption{The same as Fig.~\ref{fig:recon-unbiased-C2-r7} now for
    the Galactic region filled with the ILC values.  We see that both
    estimators are now biased to agree best near the \WMAP{} MLE value as
    we saw in Fig.~\ref{fig:recon-C2-r7-10deg-ILC}.}
  \label{fig:recon-unbiased-C2-r7-ILC}
\end{figure}

\subsection{Optimal, Unbiased $\bmath{\Cl}$ Estimator}

The general behaviour found for the naive estimator, $\Cl^e$, carry over to
the optimal, unbiased estimator, $\hat\Cl$, based on the weighted harmonic
coefficients~(\ref{eq:estimator-Cl}), as we now see.  Calculating $\hat\Cl$
for the realisations considered above we find the results in
Fig.~\ref{fig:recon-unbiased-C2-r7} for Gaussian smoothed maps.  This
figure should be compared to Fig.~\ref{fig:recon-C2-r7-10deg}.  We see that
$\hat C_2$ is nearly unbiased over the full range of true $C_2$ as expected.

The effect of smoothing when the ILC is inserted into the masked region is
shown in Fig.~\ref{fig:recon-unbiased-C2-r7-ILC}. Again we see the bias
introduced by smoothing when the two regions do not contain the same
structure.  These results are qualitatively similar to those found 
in Fig.~\ref{fig:recon-C2-r7-10deg-ILC} and the same discussion applies.

\begin{table}
  \caption{$S_{1/2}$ values for the ILC map calculated for $2\le\ell\le
    10$. The map is unprocessed, Gaussian smoothed with a $10\degr$ beam,
    or had the Galactic region filled with a Gaussian random, statistically
    isotropic sky realisation with the same power spectrum as the region
    outside this region prior to smoothing.  The values are calculated for
    the full sky and for the KQ75y7 masked sky at $\nside=128$ using a
    pixel based estimator or the optimal, unbiased $\Cl$
    estimator~(\ref{eq:estimator-Cl}) from maps at $\nside=128$ and
    $\nside=16$.  The last row refers to the map whose mask area has been
    filled with Gaussian random field whose power is consistent with power
    measured outside the mask.  }
  \label{tab:S12-ilc}
  \begin{tabular}{ccccc} \hline
    & \multicolumn{4}{c}{\boldmath $S_{1/2}\;(\mu \mathbf{K})^4$}
    \\ \cline{2-5} 
    \textbf{ILC Map} &  \textbf{Full Sky}& \textbf{Cut Sky}&
    \multicolumn{2}{c}{\textbf{Reconstructed Sky}}    \\ \cline{4-5}
    \textbf{Processing} &&  \textbf{Pixel-}
    &$\mathbf{N_{side}}$&$\mathbf{N_{side}}$\\
    &  & \textbf{based}&
    \textbf{=128} &\textbf{=16} \\ \hline
    Unsmoothed & 8835 & 1275 & 5390 & 2300 \\
    $10^\circ$ smoothing & 8835 & 1270 & 2230 & 1670 \\
    Filled with consistent & 1020 & 1290 & 1020 & 950 \\ 
    power and smoothed & &  & &  \\ 
    \hline
  \end{tabular}
\end{table}

\begin{figure}
  \includegraphics[width=0.5\textwidth]{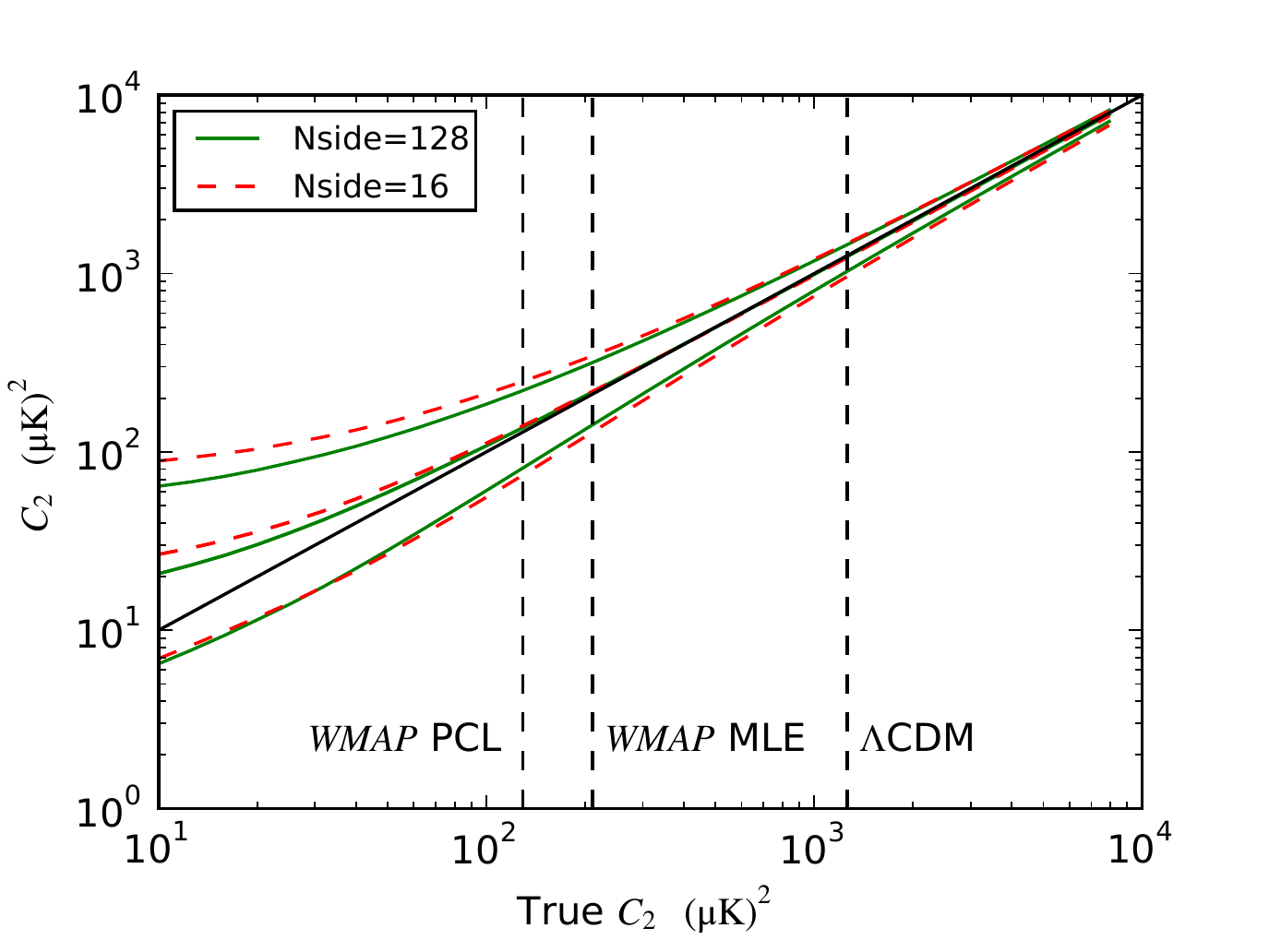}
  \caption{The same as Fig.~\ref{fig:recon-unbiased-C2-r7} now comparing
    the weighted harmonic coefficient estimator~(\ref{eq:estimator-Cl}) for
    $\nside=128$ as the green, solid lines and $\nside=16$ as the red,
    dashed lines without smoothing the map prior to reconstruction. Since
    there is no smoothing, the results do not depend on the contents of the
    masked galactic region. We see that the reconstruction without
    smoothing is unbiased for most of the $C_2$ range, however see
    Sec.~\ref{sec:nosmoothing} for a discussion of its inapplicability to
    real data.}
  \label{fig:recon-unbiased-multires-C2}
\end{figure}

\subsection{Reconstructing Without Smoothing}
\label{sec:nosmoothing}

The reconstruction of the $a_{2m}$ without smoothing showed that for
$\nside=128$ the reconstruction was unbiased
(Fig.~\ref{fig:recon-a2m-C2-100-nside-128}) but for $\nside=16$ there was a
resolution dependent bias (Fig.~\ref{fig:recon-a2m-C2-100-nside-16}).
Calculating the weighted harmonic coefficient
estimator~(\ref{eq:estimator-Cl}) from these realisations produces the
results in Fig.~\ref{fig:recon-unbiased-multires-C2}.  At first glance
these results are surprising and encouraging.  The green, solid lines for
$\nside=128$ and red, dashed lines for $\nside=16$ nearly overlap and the
central value very closely follows the true value.  This is surprising
since the $a_{2m}$ at $\nside=16$ are biased and have smaller variance than
the corresponding $\nside=128$ (see
Figs.~\ref{fig:recon-a2m-C2-100-nside-128} and
\ref{fig:recon-a2m-C2-100-nside-16}).  Even so, when combined to determine
$C_2$ these differences average out and lead to nearly identical
predictions.

Based on Fig.~\ref{fig:recon-unbiased-multires-C2} we may think we have
solved the reconstruction problem; just reconstruct using the optimal,
unbiased $\Cl$ estimator~(\ref{eq:estimator-Cl}) without smoothing!
Unfortunately we cannot draw this conclusion from the results presented
here.  Recall that the reconstructions have been performed on noise-free,
pure CMB maps.  Real maps contain noise and potentially residual, unmasked
foregrounds.  In particular uncorrected, diffuse foregrounds are known to
contaminate the low-$\ell$ reconstruction~\citep{Naselsky2008}. A careful
study of the issues faced when applying the reconstruction to real data is
beyond the scope of this work and will be reserved for future
study. However, naive application of this method to real data yields highly
biased reconstructions.

\subsection{$\bmath{S_{1/2}}$ Estimator}

The study of the $S_{1/2}$ statistic is a large project in its own right
and will not be pursued in detail here.  Our Universe as encoded in the ILC
map contains a somewhat small full-sky $S_{1/2}$ and an extremely small cut
sky $S_{1/2}$\@.  If we are to perform such a statistical study of
$S_{1/2}$ we could enforce this structure, that is, only choose skies that
have somewhat low full-sky and very low cut-sky $S_{1/2}$ values.
Alternatively we could choose from an ensemble based on the best-fitting
$\Lambda$CDM model.  In the latter case it has already been shown that the
ILC map is a rare realisation, unlikely at the $99.975\%$ level
\citep{CHSS-WMAP5}.  The assumptions made in any study will determine the
statistical questions that can be asked.  Conversely, the statistical
questions asked will implicitly contain the assumptions imposed.

In Table~\ref{tab:S12-ilc} we show the $S_{1/2}$ for the ILC map calculated
from~(\ref{eq:estimator-S12}) under various assumptions.  Note that these
values all contain the bias discussed in Sec.~\ref{sec:S12} as is standard
in the literature.  Shown in the table are the values calculated for the
full sky and for the partial sky where the KQ75y7 mask is employed to cut
out the Galactic region.  The cut-sky results are calculated using the
pixel based estimator of \spice{} and the optimal, unbiased $\Cl$
estimator~(\ref{eq:estimator-Cl}) from reconstructed maps at $\nside=128$
and $\nside=16$.  Further, the results are shown for different map
processing, including no processing (the unsmoothed entry where the map has
only been degraded as required for the reconstruction), employing a
$10\degr$ Gaussian smoothing, and filling the Galactic region with a
realisation that has the same power in each $\ell$-mode as the region
outside the mask but with the phases randomised.

The results in Table~\ref{tab:S12-ilc} are consistent with what we have
found for the $\Cl$ reconstructions.  For the unsmoothed map the full-sky
and pixel based estimators calculated at $\nside=128$ show the usual
result, the large discrepancy between the full and cut-sky values.  This
holds true for the smoothed map also.  Further, the reconstructed values
show the large discrepancy between the unsmoothed and smoothed maps (see,
for example, Figs.~\ref{fig:recon-a2m-C2-100-nside-16} and
\ref{fig:recon-a2m-C2-100-nside-16-10deg}).  We also see that the
reconstructed values are systematically larger than the pixel based
estimator showing that the reconstruction is more sensitive to leakage for
information from inside the masked region.  Finally the last line of the
table shows the expected behaviour for a map where the full sky has power
consistent with that from the cut-sky. Notice that of the cut-sky
pixel-based results are consistent with each other since information
leakage is unimportant.  The small difference between the $\nside=128$ and
$\nside=16$ reconstructions shows the residual sensitivity on resolution.

\begin{figure}
  \includegraphics[width=0.5\textwidth]{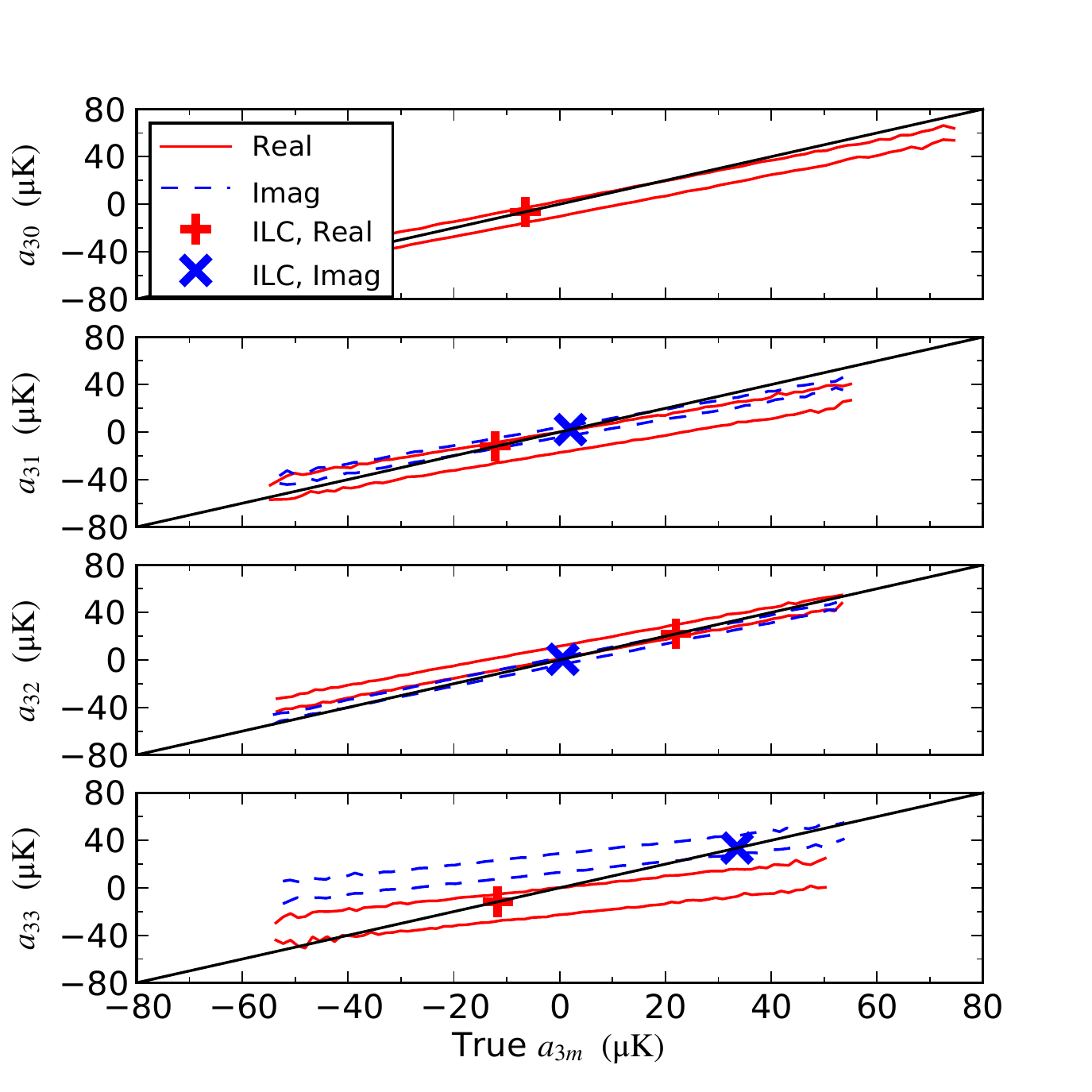}
  \caption{The $95$ and $5$ percentile lines for the $a_{3 m}$
    reconstructed from the pixels outside the KQ75y7 mask at $\nside=128$
    (and thus $\ellmax=514$) of $\Lambda$CDM realisations with
    $C_2=100\unit{(\muK)^2}$ and the masked region filled in with the ILC
    map \textit{prior} to smoothing and rescaling, as discussed in the
    text.  The red, solid lines are for the real part of the $a_{3 m}$ and
    the blue, dashed lines are for the imaginary part.  The black, solid
    line shows the expected result for a perfect reconstruction.  We
    clearly see the reconstructed $a_{3 m}$ are \textit{not} unbiased.  Now
    the bias is most prominent for $a_{33}$, the octopole mode with all its
    extrema within the Galactic plane.  This figure should be compared to
    Fig.~\ref{fig:recon-a2m-C2-100-with-ILC}.}
  \label{fig:recon-a3m-C2-100-with-ILC}
\end{figure}

\begin{figure}
  \includegraphics[width=0.5\textwidth]{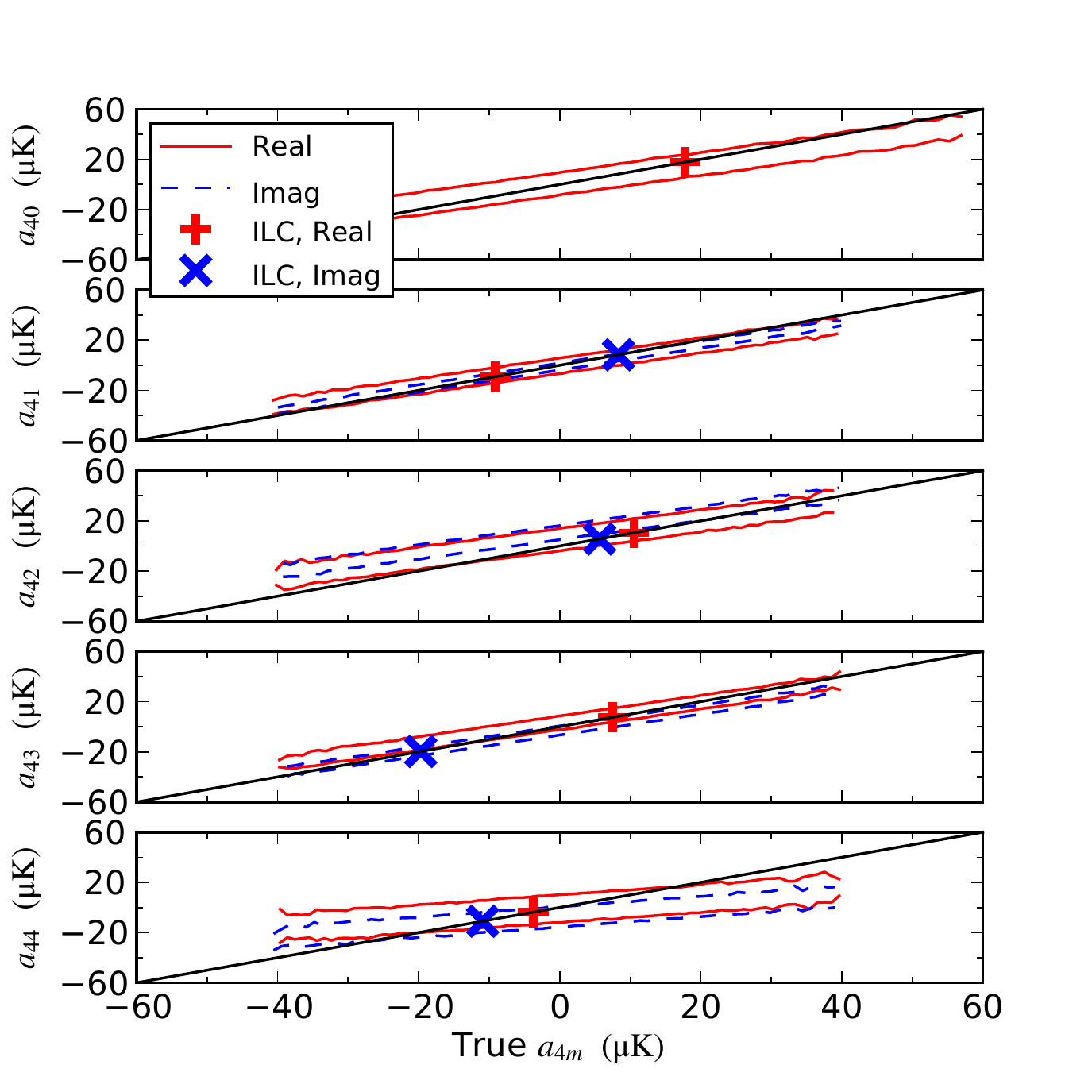}
  \caption{The same as Fig.~\ref{fig:recon-a3m-C2-100-with-ILC} now for
    $\ell=4$.}
  \label{fig:recon-a4m-C2-100-with-ILC}
\end{figure}

\subsection{Higher Multipoles}

In this work we have focused on how data handling affects the
reconstruction of the quadrupole.  The quadrupole serves as an example
of the general behaviour.  As show in
Figs.~\ref{fig:recon-a3m-C2-100-with-ILC} and
\ref{fig:recon-a4m-C2-100-with-ILC} we see the same results for $\ell=3$
and $\ell=4$.  These figures were generated from the same realisations
employed in making Fig.~\ref{fig:recon-a2m-C2-100-with-ILC}.  Again we
see that the reconstruction is biased toward the values from the ILC map.

\section{Conclusions}
\label{sec:conclusions}

It has been argued that the large-angle CMB can be reliably reconstructed
from partial-sky data and that when this is done the lack of large-angle
correlation is not significantly deviant from the expectation
\citep{Efstathiou2010}.  At first glance the argument appears sound.  The
large-angle modes extend over large fractions of the sky, thus knowing
their values on one region of the sky allows us to extrapolate them into
the masked regions.  However, in practise and under close scrutiny this
argument fails.  Implicit assumptions built in to the reconstruction
process enforce agreement between the reconstruction and the previously
constructed full sky (the ILC map in this case) through mixing of
information from inside the masked region to that outside.  Due to this the
reconstruction has no value independent of the original full-sky map.  It
neither confirms nor denies the validity of that map.

To study the large-angle CMB a choice must be made on what data to take as
a fair representation of the CMB sky One choice is to accept a cleaned,
full-sky map, such as the ILC map produced by \WMAP, to accurately
represent the primordial CMB sky.  In this case the full-sky map may be
analysed with no reconstruction required. In \cite{CHSS-WMAP5} and in this
work, however, we have taken the region outside the Galaxy as defined by
the \WMAP{} KQ75y7 mask to be a fair representation. We have shown that the
large-angle CMB can be reconstructed using unbiased estimators for the
$\alm$ and $\Cl$, however the standard approach requires processing the
original map by degrading and smoothing it.  Unfortunately it is precisely
the smoothing process that mixes the region we have taken as a fair
representation of the CMB with the region we are trying to exclude.  When
the excluded region has the same statistical properties as the region we
are including then no biases are introduced.  On the other hand, when, as
is the case with the ILC map, the properties are significantly different
the reconstruction is biased to agree with the full map.  This is not
surprising.  Through this process one is trusting the full-sky map, mixing
information from it into the rest of the sky, then reconstructing it.  This
is a circular process and is unnecessary.  If the full-sky map is already
trusted then there is no point in performing a reconstruction to produce a
poorer version of the original map.

The important point is that even \textit{in principle} reconstructing
following the standard approach leads to biased results unless the full-sky
CMB is already known.  We have shown for noise free, pure CMB maps that
smoothing mixes information and biases the results.  When applied to real
data the problems only get worse.  Encouragingly we also found that
\textit{in principle} reconstructing without smoothing leads to unbiased
results.  Unfortunately, directly applying this to real data with noise and
residual, unmasked foregrounds yields highly biased reconstructions
requiring further care to apply this method successfully to real-world CMB.

Overall the question of how to perform an unbiased reconstruction of the
full large angle CMB sky remains an interesting one.  Previous work
\citep{Bielewicz2004,Naselsky2008,Liu2009,Aurich2010} has shown that
contamination significantly affects the reconstruction of the large angle
multipole moments.  \cite{Aurich2010} studied the case most similar to that
considered in this work.  They showed that smoothing of full sky map leaks
information from the pixels not used in the reconstruction (those in a
mask) to the pixels that will be used.  In this work we have extended their
result and shown how a reconstruction such as that performed by
\cite{Efstathiou2010} is biased due to this leakage of information.  This
shows the fundamental problem with trying to reconstruct the full sky from
a partial sky.

Fortunately large-angle CMB studies are not dependent on reconstructed
full-sky maps.  The partial sky when used consistently (see
\citealt{CHSS-WMAP5}, for example) has been shown to be a robust
representation of the large scale CMB by \citet{Aurich2010} and in this
work.  Despite the fact that such an approach is suboptimal in the sense
that the inferred $\Cl$ do not have the smallest possible variance, it is
far less biased than the `optimal' $\Cl$ inferred through the
maximum-likelihood reconstruction. More robust statements about the
large-angle CMB behaviour may therefore be made with the partial sky
pixel-based $\Cl$.

We conclude that the lack of large-angle correlation, particularly on the
region of the sky outside the Galaxy, remains a matter of serious
concern. 

\section*{Acknowledgments}

We thank Devdeep Sarkar for collaboration during initial stages of this
work.  DH is supported by DOE OJI grant under contract DE-FG02-95ER40899,
and NSF under contract AST-0807564. DH and CJC are supported by NASA under
contract NNX09AC89G; DJS is supported by Deutsche Forschungsgemeinschaft
(DFG); GDS is supported by a grant from the US Department of Energy; both
GDS and CJC were supported by NASA under cooperative agreement
NNX07AG89G. This research was also supported in part by the NSF Grant
No.\ NSF PHY05-51164.  This work made extensive use of the \healpix{}
package~\citep{healpix}.  The numerical simulations were performed on the
facilities provided by the Case ITS High Performance Computing Cluster.

\bibliographystyle{mn2e_new}
\bibliography{reconstruction_bias}

\appendix

\section{Reconstructing at High Resolution}
\label{app:high-resolution}

Computationally the time and memory intensive step in reconstructing $\alm$
and $\Cl$ from our estimators~(\ref{eq:estimator-alm}) and
(\ref{eq:estimator-Cl}) is the inversion of the covariance matrix, $\mat
C$\@.  Fortunately this step only needs to be performed once for each choice
of resolution, $\nside$, and mask.

The covariance matrix is of size $\Npix\times\Npix$ where the number of
pixels is given by $12(\nside)^2$ and the size of $\mat C$ scales as
$(\nside)^4$.  An increase in resolution by one step, $\nside\rightarrow
2\nside$, increases the size of $\mat C$ by a factor of $16$\@. Working
with cut skies does not appreciably reduce this, even the largest mask,
KQ75y7, only cuts out $25$--$30$ percent of the pixels.  Resolutions of
$\nside=32$ or perhaps even $\nside=64$ are attainable on a desktop
computer.  Fortunately we never need to store the full $\Cinv$ and can
calculate the elements of $\mat C$ as required instead of storing them.

In our estimators all the matrices that we encounter, except for $\Cinv$,
are of size $N_\ell\times\Npix$ or smaller. Here
$N_\ell=(\ellrecon+1)^2$. Even for $\nside=512$ and $\ellrecon=10$ these
matrices only require about $3\unit{GB}$ of storage at double precision.
Further we see that only the matrix
\begin{equation}
  \mat M\equiv\Cinv\mat Y
  \label{eq:mat-M}
\end{equation}
is ever required (see Eqs.~\ref{eq:estimator-alm} and
\ref{eq:covariance-alm}).

To compute $\mat M$ we note that it satisfies the set of linear equations
\begin{equation}
  \mat C\mat M = \mat C\Cinv\mat Y = \mat Y.
\end{equation}
Solving such a set of equations is a standard problem in computational
linear algebra.  A covariance matrix is symmetric and positive-definite so
it may be factored with a Cholesky decomposition \citep{numerical-recipes}
\begin{equation}
  \mat C = \mat L\transpose{\mat L},
\end{equation}
where $\mat L$ is a lower triangular matrix.  Our problem then becomes
solving 
\begin{equation}
  \mat L(\transpose{\mat L}\mat M) \equiv \mat L\mat z=\mat Y.
\end{equation}
This can be solved in two steps using backward substitution on $\mat
L\mat z=\mat Y$ to find $\mat z$ followed by forward substitution on
$\transpose{\mat L}\mat M=\mat z$ to find $\mat M$.\@

At this point we are left with computing $\mat L$\@.  Approximately half of
this matrix is zero so only half of it needs to be stored (of course the
same is true of $\mat C$ since it is symmetric).  Unfortunately this cannot
be further reduced and this provides the limiting factor in determining the
resolution at which we can work.  For $\nside=128$ and $\ellrecon=10$ the
matrix $\mat L$ is approximately $70\unit{GB}$ in size.  Improving
resolution to $\nside=256$ increases the required storage to over
$1\unit{TB}$.\@ This is what has limited our work to $\nside=128$.
Straight forward, numerically stable algorithms exist for calculating $\mat
L$ (see \citealt{numerical-recipes}, for example).  Though this is a time
consuming step once $\mat M$ is calculated the rest follows quickly.

\bsp

\label{lastpage}

\end{document}